\documentclass[twoside,compsoc]{IEEEtran}
\usepackage{makecell}
\usepackage{hyperref}
\usepackage{array}
\usepackage{graphicx,amssymb,amsmath}
\usepackage{multicol}
\usepackage[noadjust]{cite}
\usepackage{setspace}
\usepackage{subfigure}
\usepackage{float}
\usepackage {url}
\usepackage{stfloats}
\usepackage{amsthm,pifont}
\usepackage{flushend}
\usepackage[table]{xcolor}
\usepackage{multirow}
\usepackage{caption}
\usepackage{rotating}
\usepackage{cases,subeqnarray}
\usepackage{bm,multirow,bigstrut}
\usepackage{textcomp}
\usepackage{latexsym,bm}
\usepackage{booktabs}
\usepackage{xcolor}
\usepackage{mathtools}
\usepackage{dsfont}
\usepackage{extarrows}
\usepackage{epsfig}
\usepackage{epsfig}
\usepackage{epstopdf}
\usepackage{colortbl}
\usepackage{algpseudocode}
\usepackage{algorithmicx,algorithm}

\theoremstyle{plain}

\theoremstyle{plain}

\usepackage{amsmath}

\definecolor{Gray}{gray}{0.85}
\begin{document}
\title{JPPO++: Joint Power and Denoising-inspired Prompt Optimization for Mobile LLM Services}

\author{Feiran You, Hongyang Du, Kaibin Huang,~\IEEEmembership{Fellow,~IEEE}, and Abbas~Jamalipour,~\IEEEmembership{Fellow,~IEEE}
\thanks{F. You, H. Du, and K. Huang are with the Department of Electrical and Electronic Engineering, University of Hong Kong, Pok Fu Lam, Hong Kong SAR, China (email: fryou@eee.hku.hk, duhy@eee.hku.hk, huangkb@hku.hk).}
\thanks{A. Jamalipour is with the School of Electrical and Computer Engineering, University of Sydney, Sydney, Australia (email: a.jamalipour@ieee.org).}
\thanks{The conference version of this work has been accepted by the IEEE International Conference on Communications (ICC) 2025~\cite{you2024jppojointpowerprompt}.}
}
\maketitle

\vspace{-1cm}
\begin{abstract}
Large Language Models (LLMs) are increasingly integrated into mobile services over wireless networks to support complex user requests. This trend has led to longer prompts, which improve LLMs' performance but increase data transmission costs and require more processing time, thereby reducing overall system efficiency and negatively impacting user experience. 
To address these challenges, we propose Joint Prompt and Power Optimization (JPPO), a framework that jointly optimizes prompt compression and wireless transmission power for mobile LLM services. JPPO leverages a Small Language Model (SLM) deployed at edge devices to perform lightweight prompt compression, reducing communication load before transmission to the cloud-based LLM. A Deep Reinforcement Learning (DRL) agent dynamically adjusts both the compression ratio and transmission power based on network conditions and service constraints, aiming to minimize service time while preserving response fidelity. 
We further extend the framework to JPPO++, which introduces a denoising-inspired compression scheme. This design performs iterative prompt refinement by progressively removing less informative tokens, allowing for more aggressive yet controlled compression. Experimental results show that JPPO++ reduces service time by $17\%$ compared to the no-compression baseline while maintaining output quality. Under compression-prioritized settings, a reduction of up to $16\times$ in prompt length can be achieved with an acceptable loss in accuracy. Specifically, JPPO with a $16\times$ ratio reduces total service time by approximately $42.3\%$, and JPPO++ further improves this reduction to $46.5\%$.
\end{abstract}
\IEEEpeerreviewmaketitle

\section{Introduction}
The rapid advancement of Artificial Intelligence (AI) has positioned it as a key enabler for Sixth Generation (6G) communication systems~\cite{you2024jppojointpowerprompt}, supporting a broad range of intelligent services. Among AI technologies, Large Language Models (LLMs) have become central to applications such as question answering, code generation, and real-time dialogue~\cite{he2024large, boateng2025survey}. 
Their ability to interpret and generate context-aware information has significantly advanced natural language processing, enabling a variety of services that are accessed through both cloud and edge environments~\cite{yi2025edgemoe, qu2025mobile}. As users increasingly access these services via mobile devices, wireless networks have become a critical infrastructure for delivering LLM-powered intelligence~\cite{ren2024industrial}. 

Unlike traditional network-aided services that primarily involve structured data transmission, LLM-based applications impose significantly higher communication demands due to the complexity and variability of natural language prompts and generated responses. 
Moreover, LLM services often require real-time interaction and are sensitive to prompt transmission delays, presenting new challenges for wireless systems, especially under constraints such as limited transmit power, fluctuating channel conditions, and restricted edge computing resources. 
These challenges are further amplified in mobile edge scenarios, where users interact with cloud-hosted LLMs via edge devices by sending prompts and receiving responses over wireless links~\cite{10669603}. This mode of service places substantial pressure on the wireless infrastructure, exposing a mismatch between the LLM service requirements and available communications and computing resources. Addressing this mismatch requires new system-level optimization frameworks that adaptively manage transmission and processing strategies to maintain service quality under dynamic network conditions~\cite{min2023recent, jiang2024large}.

Among the various factors contributing to the system load, the structure and length of the input prompt play a central role. In LLMs, a prompt is first tokenized into a sequence of subword units, known as tokens~\cite{patil2024review}. These tokens are embedded and passed through a transformer architecture comprising multiple layers of self-attention and feed-forward networks~\cite{9996581}, which model contextual dependencies and generate the final output. This computation scales with prompt length, as each token attends to all previous tokens, resulting in quadratic time and memory complexity. Furthermore, recent prompting strategies such as In-Context Learning (ICL)~\cite{li2024long} and Chain-of-Thought (CoT)~\cite{wei2022chain} have pushed LLMs toward more advanced reasoning, but often at the cost of significantly longer prompts~\cite{10705427}. In practical scenarios, users may upload entire documents along with queries, such as research papers or legal texts containing thousands of words, to elicit accurate and context-rich responses. While these techniques improve LLM service quality, they commonly result in prompt lengths reaching tens of thousands of tokens~\cite{xue2024repeat}. This introduces a fundamental trade-off: longer, more informative prompts enhance output quality but impose heavy burdens on wireless transmission and edge-side inference.

\begin{figure}[t]
\centering
\includegraphics[width=0.496\textwidth]{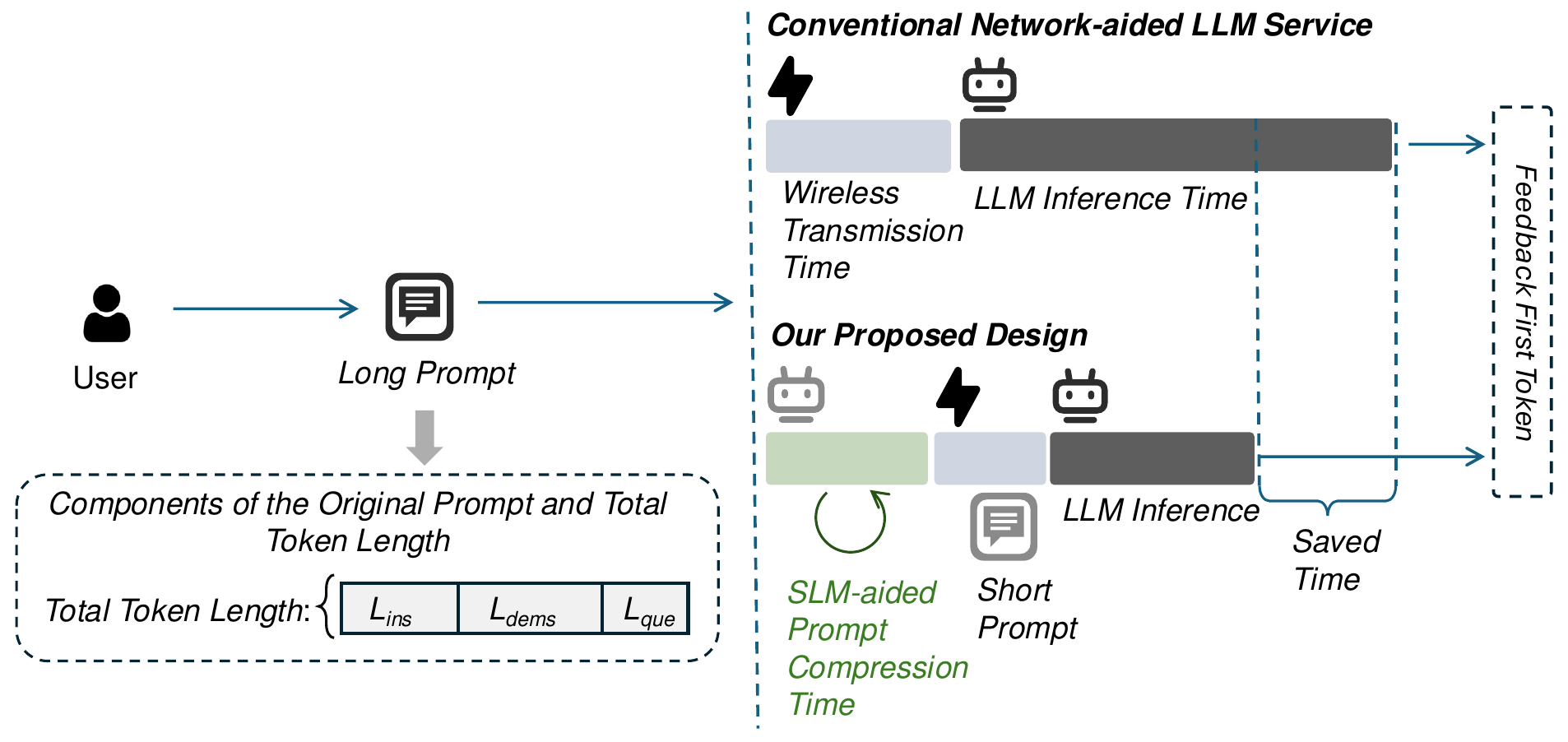}
\caption{Time consumption comparison of first token generation between conventional network-aided LLM inference service architecture and our design under long user prompts.}
\label{figure1}
\end{figure}
Prior research has approached these challenges from both the model and network perspectives. From the model side, the authors in~\cite{jiang2023llmlingua} introduced LLMLingua, a coarse-to-fine prompt compression method that demonstrates significant potential for compressing LLM prompts while preserving their semantic integrity. LLMLingua utilizes a Small Language Model (SLM) for compression, which can be aligned with the target LLM through instruction tuning. While this approach shows promise for reducing communication overhead, it does not fully account for the challenges posed by wireless transmission. 
From the network side, LLM-Slice~\cite{liu2024llm} introduces a wireless slicing mechanism that reserves dedicated network resources for LLM services, thereby reducing response latency. Yet this method focuses purely on communication resource allocation and does not incorporate prompt-level adaptation tailored to individual service requirements. These limitations highlight the need for a unified solution to address two key questions:
\begin{itemize}
\item {\textbf{Q1)}} How to achieve adjustable prompt compression without significantly degrading LLM inference performance and service quality?
\item {\textbf{Q2)}} How to jointly allocate wireless resources, particularly transmission power, to meet the latency and energy constraints of network-aided LLM services?
\end{itemize}
For {\textbf{Q1}}, natural language processing methods offer potential solutions for prompt compression. However, traditional NLP compression techniques, such as text summarization or keyword extraction, often fail to capture the complex reasoning patterns and task-specific requirements embedded in LLM prompts, leading to degraded inference performance. Alternative AI-based solutions, including large autoencoder models or task-specific compression networks, typically demand substantial computational resources and introduce additional latency at the user side, rendering them unsuitable for wireless scenarios with constrained devices. 
Motivated by LLMLingua~\cite{jiang2023llmlingua}, we consider SLMs as a practical solution for prompt compression to achieve resource-efficient deployment. SLMs can be easily integrated into user terminals and offer sufficient semantic understanding to retain task-critical content during compression. They have also been shown to be effective in enabling transformer-based inference in edge applications~\cite{10745806}.

Furthermore, inspired by the diffusion models, especially the Denoising Diffusion Probabilistic Models (DDPM)~\cite{ho2020denoising}, we propose an iterative prompt compression algorithm to complement direct SLM-based compression. In this approach, the original long prompt is treated as a ``noisy'' input, and the refined compressed version represents the ``denoised'' state. Analogous to how diffusion models progressively remove noise from images, our method performs multi-stage prompt refinement, gradually filtering out redundant or non-essential content. This design mitigates the risk of semantic loss that can occur when compressing lengthy prompts in a single step.
This principle is similar to how humans revise long-form text: reducing a complex document by $90\%$ in one attempt often results in missing key points, while multiple rounds of revision, each removing a manageable portion, allow for better content preservation and clarity. Likewise, a $16\times$ compression can be achieved through four iterations of $2\times$ compression, enabling the model to operate on smaller, semantically coherent segments at each stage. This controlled process enhances the likelihood of retaining task-critical information while maintaining high compression. Despite its iterative structure, the method introduces minimal computational overhead due to the efficiency of SLMs. Empirically, each compression round adds only about $2\%$ to the total LLM inference time, making it well-suited for latency-sensitive wireless applications. 

Building upon this insight and for addressing the {\textbf{Q2}}, we propose Joint Power and Denoising-inspired Prompt Optimization (JPPO++), a framework that combines SLM-based prompt compression with wireless power allocation optimization, as illustrated in Fig.~\ref{figure1}. JPPO++ captures the trade-off between compression ratio and wireless resource consumption using Deep Reinforcement Learning (DRL), adapting to both channel conditions. The contributions of this paper are summarized as
\begin{itemize}
\item We utilize a small, aligned language model as a prompt compressor in JPPO++ for network-aided LLM services. The SLM efficiently captures essential meaning while significantly reducing the length of the prompt without requiring training at the transmitter. Additionally, we design a denoising-inspired prompt compression mechanism that performs compression through iterative refinement steps, further improving the system's overall performance with marginal additional computational overhead.
\item We formulate a joint optimization problem that captures the trade-offs among wireless transmission efficiency and LLM service quality. The objective is to maximize end-to-end Quality of Service (QoS) under energy and delay constraints, where the compression ratio and transmission power serve as coupled decision variables.
\item To solve this dynamic optimization problem, we employ a Deep Reinforcement Learning (DRL) framework. The DRL agent learns adaptive compression and power strategies based on real-time network conditions and prompt characteristics. This enables JPPO++ to outperform static baselines by reducing communication cost and improving responsiveness in resource-constrained wireless environments.
\end{itemize}
The remainder of this paper is organized as follows: Section 2 discusses related work in LLM deployment, prompt optimization, and wireless resource allocation. Section 3 presents our system model for wireless network-aided LLM services and introduces the novel denoising-inspired prompt compression method. Section 4 details the JPPO++ framework, including its mathematical foundations and implementation approach. Section 5 presents comprehensive numerical results and performance analysis. Finally, Section 6 concludes the paper with a summary of our key findings.
A list of mathematical symbols frequently used in this paper is shown in Table~\ref{table1xxx}.
\begin{table}[t]
\caption{Mathematical Notations}
\vspace{-0.2cm}
\centering
\label{table1xxx}
\renewcommand{\arraystretch}{1}
\begin{tabular}{m{2.2cm}|m{5.8cm}}
\toprule
\textbf{Notation} & \textbf{Description} \\
\midrule
${\textit{\textbf{x}}}_{\text{ins}}, {\textit{\textbf{x}}}_{\text{dems}}, {\textit{\textbf{x}}}_{\text{que}}$ & The instruction component, the demonstrations or examples, and the specific question or task that consisted in an original prompt $\textit{\textbf{x}}$ \\
$\hat{\textit{\textbf{x}}}$ & The compressed prompt\\ 
$\mathcal{L}_x, \mathcal{L}_{\hat{x}}$ & The total token length of the original
prompt and  the token length of the compressed prompt\\
$\mathbf{f_1} ,\mathbf{f_2}, \mathbf{f_3}$ & The three
essential aspects of semantic preservation during information transmission to evaluate the quality of prompt compression \\
$\phi_1, \phi_2, \phi_3 $ & The weight factors determining the relative importance of each fidelity component \\
$\alpha(t) $ & The compression ratio at any time $t$ \\
$\psi_\kappa$ & The target compression ratio\\
$\sigma(t)$  & A monotonically increasing scheduling function\\
$\beta(i)$ & The compression ratio between consecutive steps \\
$E_e$, $E_t$ & The encoding energy and transmission energy \\
$R$ & The transmission rate  \\
$w$ & The bandwidth of the offloading link between the user and DCO \\
$\gamma$ & The Signal-to-Noise Ratio (SNR) \\
$\omega$ & The path-loss exponent \\
$\lambda^2$ & The Gaussian noise term in the Additive White Gaussian Noise (AWGN) channel \\
$T$ & The total time consumption \\
$t_{\text{e}}^{{\rm SLM}}$, $t_{\text{e}}^{{\rm LLM}}$, $t_{\text{t}}$ & The encoding delay in both SLM and LLM, and transmission
delay \\
$\varphi$ & The Probability Density Function (PDF) expression for wireless channel fading\\
$P_{\text{T}}$ & The transmit power\\
$\eta$ & BEP \\
$P_{\text{th}}$ & The maximum allowable power consumption \\
$T_{\text{th}}$ & The maximum tolerable end-to-end latency \\
$\mathbf{f}_{\text{th}}$ & The minimum required fidelity \\
\bottomrule
\end{tabular}
\end{table}

\section{Related Work}
This section discusses several related works about LLMs, prompt compression, and wireless network management.

\subsection{Large Language Model Services}
The integration of LLMs into wireless network-aided services has enabled a range of intelligent applications, including real-time decision-making and natural language understanding~\cite{10582827}. However, this trend imposes increasing demands on both computation and communication resources due to the growing size of LLMs and the complexity of their input prompts. To address computational bottlenecks, several recent efforts have focused on optimizing LLM inference at the network edge. For example, the authors in~\cite{10571127} formulate a scheduling problem that jointly allocates computing and transmission resources for transformer-based LLMs, maximizing throughput under edge constraints. Similarly, the authors in~\cite{10759588} propose an edge inference framework that combines model quantization and batching, and develop the Optimal Tree-search with Generalized Assignment Heuristics (OT-GAH) algorithm for efficient scheduling on resource-limited devices. These studies highlight the importance of inference-side optimization, yet they operate under an implicit assumption that the input prompt is already available. In practice, especially in cloud-based LLM deployments accessed via wireless networks, uploading long prompts often accounts for a substantial portion of the total latency. This issue is exacerbated by recent prompting strategies, such as ICL and CoT reasoning, which require longer, more structured inputs to fully exploit model capabilities. As prompt length grows, communication costs become a dominant factor in end-to-end performance, particularly in bandwidth-limited or energy-constrained environments. While the authors in~\cite{10599304} offer a broad vision for aligning foundation models with the needs of wireless systems and discuss emerging Large Multi-modal Models (LMMs), they do not directly address prompt transmission overhead or latency. This gap highlights a critical insight: without coordinated control of both prompt input and transmission behavior, LLM services over wireless networks will remain bottlenecked by front-end communication delays. Thus, prompt compression and adaptive resource control must be treated as essential optimization targets, jointly considered alongside inference design to support scalable, low-latency LLM services in next-generation networks.

\subsection{Prompt Compression}
Prompt compression has emerged as a crucial technique for mitigating the computing and communication overhead introduced by increasingly lengthy prompt inputs in LLM services. However, existing compression methods face significant challenges, including limited compression efficiency, reliance on fine-tuning or task-specific heuristics, and the risk of degrading LLM performance due to loss of critical contextual information~\cite{wang2024adapting}. 
These limitations are particularly critical in wireless environments, where bandwidth and energy constraints amplify the impact of prompt length. Moreover, conventional interference management techniques become less effective when signal and interference power levels converge, further degrading system-level performance~\cite{10646365}. Several recent works have attempted to improve prompt processing efficiency. VR-CMC~\cite{10236483} introduces variable-rate prompts for cross-modal compression, allowing data to be represented at multiple levels of granularity. PCRL~\cite{10535182} formulates prompt editing as a discrete decision process and applies reinforcement learning to learn efficient compression policies. Furthermore, the authors in~\cite{wang2024adapting} propose a framework that integrates summarization, soft prompt compression, and utility-aware mechanisms to reduce context processing load while maintaining model utility across tasks. While these approaches offer meaningful gains in compression quality and model efficiency, they are primarily designed without considering the dynamics of wireless systems. Crucially, none of these methods consider the joint effects of prompt compression and wireless transmission control. In bandwidth and energy-constrained settings, compression alone reduces data volume but cannot guarantee reliable or timely delivery. Conversely, adaptive power allocation improves link quality but does not reduce the underlying transmission load. These two levers—prompt compression and wireless resource control—are inherently complementary. Their joint optimization is crucial for enabling efficient, low-latency LLM services over wireless networks, particularly under stringent QoS and energy constraints.

\subsection{Wireless Network Management}
Extensive research in wireless network management has led to effective solutions for power allocation and joint optimization under resource constraints. For example, the authors in~\cite{10679202} propose the Delayed-Interaction Collaborative-Learning Independent-Decision Multi-Agent DRL (DICLID-MADRL) algorithm, where access points independently optimize user selection and power configuration using only local observations, improving global energy efficiency through decentralized multi-agent reinforcement learning. Beyond power control, recent studies have addressed more complex joint optimization problems. In~\cite{10679218}, a block coordinate descent and successive convex approximation framework is applied to jointly optimize receive beamforming, power allocation, HAPS positioning, and computation resource distribution. Similarly, the authors in~\cite{10545547} employ a graph neural network-based method to jointly solve power control and spectrum allocation in multi-user interference scenarios with shared channels.
These works demonstrate the potential of learning-based and algorithmic strategies for managing wireless resources in complex environments. However, they do not account for the unique characteristics of network-aided LLM services, where large, dynamic prompt inputs and strict latency constraints fundamentally alter the nature of the optimization problem. In particular, the interplay between input compression and transmission power remains unaddressed. Bridging this gap requires extending joint optimization frameworks to explicitly incorporate the semantic structure and variability of LLM inputs alongside traditional physical-layer resource control.

In summary, while prior studies have made significant progress in both prompt compression and wireless resource management, they treat these aspects in isolation. Our work departs from this separation by introducing JPPO++, which enables efficient and scalable LLM service delivery in resource-constrained wireless networks.

\section{System Model}
In this section, we present the system model of wireless network-aided mobile LLM services, the SLM-based prompt compression method, and the wireless transmission model with constraints on power consumption and service delay. 

\begin{figure}[t]
\centering\includegraphics[width=0.45\textwidth]{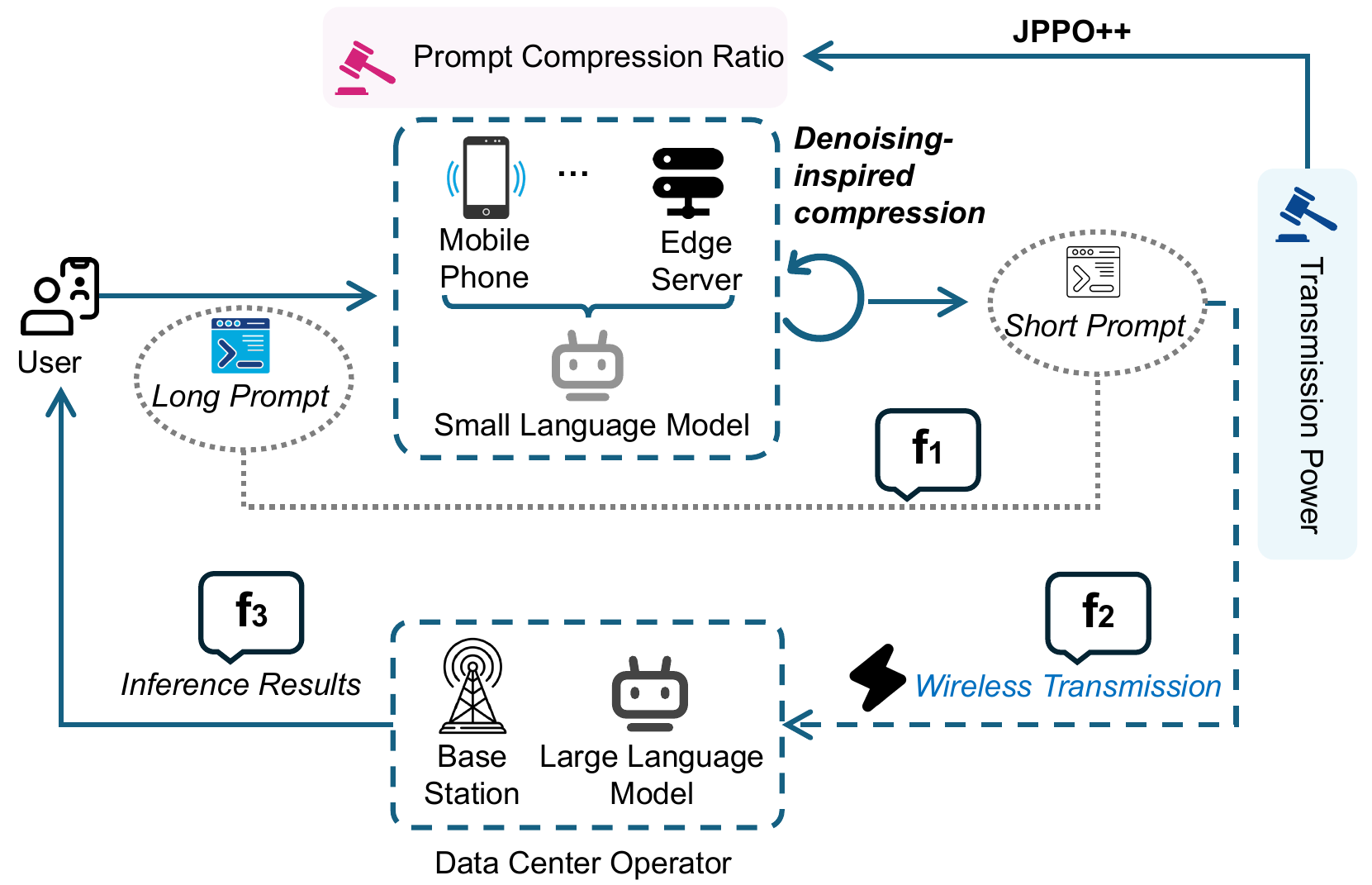}
\caption{System model of wireless network-aided mobile LLM services and overview of our proposed JPPO++, where user-generated long prompts are first compressed through SLM-based edge computing, then transmitted with optimized power allocation via wireless networks to the LLM server, and finally inference results are returned to users. The illustration of the three components in the overall fidelity metric is also demonstrated: $\mathbf{f_1}$ represents the representation accuracy, $\mathbf{f_2}$ denotes the transmission completeness, and  $\mathbf{f_3}$ is the understanding accuracy. }

\label{figuresystem}
\end{figure}

\subsection{Wireless Network-aided LLM Services}
We consider a heterogeneous wireless network where a Data Center Operator (DCO) provides LLM inference services to $N$ users with diverse task requirements, e.g., prompts.
As illustrated in Fig.~\ref{figuresystem}, our proposed framework consists of three key components: an SLM agent deployed at user devices or edge servers for prompt compression, a target LLM for inference service, and a JPPO++ scheme for reliable wireless transmission. On the user side, the SLM agent leverages its semantic understanding capability to compress prompts while preserving task-critical information. The compressed prompts are then transmitted through wireless channels with jointly optimized power allocation and finally processed by the target LLM for inference. This framework adaptively adjusts both compression ratios and transmission power based on channel conditions and prompt characteristics to achieve high LLM service quality.

\subsection{Prompt Compression}
To efficiently reduce prompt sizes while preserving semantic integrity, we adopt a coarse-to-fine compression approach in SLMs. Our compression method is designed to achieve two primary objectives: preserving critical information in the prompt while ensuring the recoverability of the original semantic meaning, and enabling flexible compression ratios that can be dynamically adjusted together with communication resources.

Let us formally define the prompt structure and compression process. Consider an original prompt ${\textit{\textbf{x}}}$ that consists of three components as:
\begin{equation}
{\textit{\textbf{x}}} = \left({\textit{\textbf{x}}}_{\text{ins}}, {\textit{\textbf{x}}}_{\text{dems}}, {\textit{\textbf{x}}}_{\text{que}}\right),
\end{equation}
where ${\textit{\textbf{x}}}_{\text{ins}}$ represents the instruction component, ${\textit{\textbf{x}}}_{\text{dems}}$ denotes the demonstrations or examples, and ${\textit{\textbf{x}}}_{\text{que}}$ contains the specific question or task. As depicted in Fig. 1, the token length of each component is denoted by $\mathcal{L}_{\text{ins}}$, $\mathcal{L}_{\text{dems}}$, and $\mathcal{L}_{\text{que}}$, respectively. The total token length $\mathcal{L}_x$ of the original prompt is given by:
\begin{equation}
\mathcal{L}_x = \mathcal{L}_{\text{ins}} + \mathcal{L}_{\text{dems}} + \mathcal{L}_{\text{que}}.
\end{equation}
Our SLM-based compression mechanism generates a compressed prompt $\hat{\textit{\textbf{x}}}$ with length $\mathcal{L}_{\hat{x}}$. The compression ratio $\alpha$ is defined as:
\begin{equation}
\alpha = \frac{\mathcal{L}_{\hat{x}}}{\mathcal{L}_x}, \quad \alpha \in [0, 1]
\end{equation}
where $\alpha = 1$ indicates no compression and smaller $\alpha$ represent higher compression rates.

We evaluate the quality of prompt compression using a composite fidelity metric  $\mathbf{f}$ that captures three complementary aspects: semantic preservation, transmission integrity, and interpretability by the downstream LLM. These three components jointly reflect the end-to-end effectiveness of delivering and using compressed prompts in wireless network-aided LLM services, as shown in Fig.~\ref{figuresystem}:
\begin{itemize}
\item \textit{Representation accuracy} ($f_1$): measures how accurately the compressed prompt preserves the semantic content of the original prompt. It reflects the degree to which key information is retained during the compression process, independent of the channel effects. This component is computed based on token-level overlap between the original prompt $\textit{\textbf{x}}$ and the compressed prompt $\hat{\textit{\textbf{x}}}$ as:
\begin{equation}
f_1 = \frac{|\hat{\textit{\textbf{x}}} \cap \textit{\textbf{x}}|}{|\textit{\textbf{x}}|},
\end{equation}
where $|\hat{\textit{\textbf{x}}} \cap \textit{\textbf{x}}|$ denotes the number of overlapping tokens. This metric serves as a direct measure of semantic fidelity after the compression mechanism.

\item \textit{Transmission completeness} ($f_2$): evaluates the fraction of the compressed prompt that is successfully transmitted under practical wireless channel conditions. It captures the effect of both prompt length reduction and bit-level errors due to signal degradation. Formally, $f_2$ is defined as:
\begin{equation}
f_2 = \frac{|\hat{\textit{\textbf{x}}}|}{|\textit{\textbf{x}}|} \cdot \left(1 - \text{BEP}(\text{SNR})\right),
\end{equation}
where $\text{BEP}(\cdot)$ denotes the bit error probability as a function of the Signal-to-Noise Ratio (SNR) ratio. This term models the effective information retained after wireless transmission, taking into account real-world physical-layer impairments.

\item \textit{Understanding accuracy} ($f_3$): assesses the utility of the compressed prompt from the LLM’s perspective. It quantifies how well the transmitted prompt enables the LLM to generate an output that aligns with the expected response to the original uncompressed input. This is measured via token-level overlap between the LLM's output $\hat{\textit{\textbf{y}}}$ and the reference original response $\textit{\textbf{y}}$ as:
\begin{equation}
f_3 = \frac{|\hat{\textit{\textbf{y}}} \cap \textit{\textbf{y}}|}{|\textit{\textbf{y}}|}.
\end{equation}
This component captures the ultimate effectiveness of the entire prompt compression and transmission pipeline in supporting the downstream task.

\end{itemize}

Together, these three fidelity components form a holistic evaluation framework. $f_1$ focuses on semantic preservation during compression, $f_2$ captures the degradation during transmission, and $f_3$ reflects the downstream interpretability and task utility. The overall fidelity metric $\mathbf{f}$ can be defined as a weighted sum of these components as:
\begin{equation}
\mathbf{f} = \phi_1\mathbf{f_1} + \phi_2\mathbf{f_2} + \phi_3\mathbf{f_3},
\end{equation}
where $\phi_1$, $\phi_2$, and $\phi_3$ are weight factors determining the relative importance of each fidelity component. Additionally, these weights can be adjusted based on LLM application requirements and QoS priorities. 

\subsection{Denoising-inspired Prompt Compression}

We propose a framework for controlling the compression ratio across multiple steps, inspired by the denoising process in DDPM~\cite{ho2020denoising}.

\subsubsection{Motivation}
\begin{figure*}[t]
\centering\includegraphics[width=0.95\textwidth]{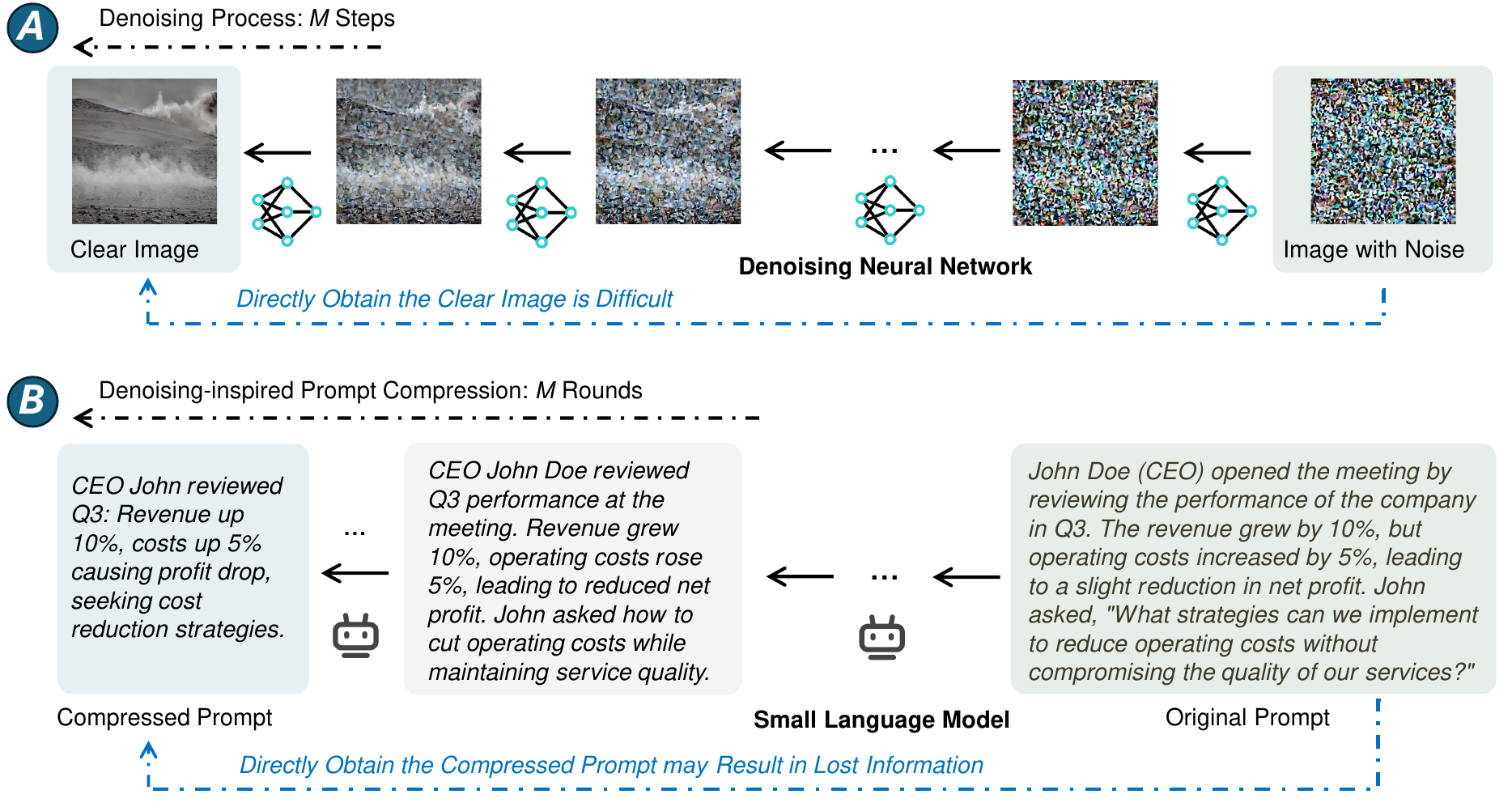}
\vspace{0.3cm}
\caption{The motivation of our proposed denoising-inspired prompt compression method. Part A illustrates the process of generating images using DDPM through a denoising process. Rather than directly generating a clear image, the denoising neural network focuses on predicting noise one step at a time, using the image from the previous step to generate the current one. Part B illustrates the denoising-inspired prompt compression. Instead of directly obtaining a fully compressed prompt, SLM can also perform gradual prompt compression to reduce the degree of information that needs to be compressed in each round of processing, minimizing information loss.}
\label{compareshow}
\end{figure*}
As shown in Part A of Fig.~\ref{compareshow}, in DDPM's denoising process, the denoising network predicts the noise between the current state and the previous state, gradually removing noise through this process until a clean image is obtained. Compared to many AI algorithms that directly generate images, DDPM's approach reduces training difficulty and improves image generation performance.

Similarly, when we use SLM for text compression, the SLM continuously predicts the next token based on the input text to ultimately obtain the compressed prompt, as shown in Part B of Fig.~\ref{compareshow}. If we set a very high compression ratio all at once, the SLM might lose crucial information. However, if we consider denoising-inspired prompt compression, where the compressed prompt is viewed as a clean image and the original long prompt is seen as a noisy state, we can enable the SLM to perform gradual compression of the prompt.
During each compression step, we can select a relatively larger compression ratio, reducing the degree of compression, and achieve the same compression effect through multiple iterations. This way, during each compression step, the SLM needs to consider relatively less information to reduce, resulting in a lower probability of losing important information, as shown in Fig.~\ref{examplefa} in Section~\ref{nums}.

Although the iterative nature introduces additional SLM processing time, the system maintains its efficiency due to SLM's swift operation. Experiments in Section~\ref{nums} show that each compression round only adds approximately $2\%$ to the total LLM inference time, while the optimized compression scheme can reduce subsequent LLM inference delay by around $40\%$.

\subsubsection{Compression Ratios Scheduling}
JPPO++ allows flexible scheduling of compression ratios while maintaining a deterministic path toward the target compression rate. Given the target compression ratio $\psi_\kappa = 1/\kappa$, e.g., $\psi = 16$ for $16\times$ compression, we define a compression progress variable as $t \in [0,1]$, where $t = 0$ represents the original text and $t = 1$ represents the fully compressed text. The compression ratio at any time $t$ is given by:
\begin{equation}
\alpha(t) = \psi_\kappa^{-\sigma(t)},
\end{equation}
where $\sigma(t)$ is a monotonically increasing scheduling function with $\sigma(0) = 0$ and $\sigma(1) = 1$. 

Considering a sequence of $M$ compression steps, we can discretize $t$ into $\{t_0, t_1, ..., t_M\}$, where $t_0 = 0$ and $t_M = 1$. The compression ratio between consecutive steps is:
\begin{equation}
\beta(i) = \frac{\alpha(t_i)}{\alpha(t_{i-1})} = \psi_\kappa^{-\left(\sigma(t_i) - \sigma(t_{i-1})\right)}.
\end{equation}
To enable flexible management of the compression process, we employ different scheduling methods tailored to various input prompt types, thereby allowing for efficient management of the compression process. 
The behavior of the compression process can be controlled through different choices of $\sigma(t)$:

\begin{itemize}
\item {\textit{The linear schedule}} is defined as:
\begin{equation}
\sigma(t) = t.
\end{equation}
This schedule applies a uniform compression rate throughout the entire process, ensuring consistent behavior across all steps, which is straightforward and computationally efficient.

\item {\textit{The cosine schedule}} is defined as:
\begin{equation}
\sigma(t) = \frac{1 - \cos(\pi t)}{2}.
\end{equation}
This schedule introduces smooth transitions at the beginning and end of the process. By tapering the compression rate during these stages, it minimizes abrupt changes, making it ideal for applications where gradual adjustments are critical to maintaining output quality.

\item {\textit{The quadratic schedule}} is defined as:
\begin{equation}
\sigma(t) = t^2.
\end{equation}
This schedule concentrates the majority of compression in the later stages of the process, leaving earlier stages relatively unaffected, which would be effective for tasks requiring precise compression towards the final steps.
\end{itemize}

Different types of long prompts may require different optimal compression schedules to ensure efficiency. 
For a uniform $M$-step compression process with linear scheduling, the discrete time steps are given by $t_i = \frac{i}{M}$, where $i \in \{0,1,\ldots, M \}$, with each step applying a constant compression ratio of $\psi_\kappa^{-1/M}$. The total compression at any time $t$ satisfies $\alpha(t) \cdot \alpha\left(0\right) = \psi_\kappa^{-\sigma(t)}$. This formulation ensures that the initial state maintains the original length (i.e., $\alpha\left(0\right) = 1$), the final state achieves the target compression (i.e., $\alpha(1) = \frac{1}{\psi_\kappa}$), and the compression path remains continuously differentiable when using smooth scheduling functions. This framework offers a principled approach to designing multi-step compression strategies, allowing for precise control over the compression ratio at each step. 

\subsection{Energy Consumption}
The total energy consumption $E$ in the one-shot LLM service request process for each user consists of two components: encoding energy $E_e$ and transmission energy $E_t$, as is given by:
\begin{equation}
E\left(\kappa, P_T\right) = E_e\left(\kappa\right) + E_t\left(\kappa, P_T\right),
\end{equation}
where $P_{\text{T}}$ is the transmit power. The encoding energy consumption $E_e$, which represents the energy used by the SLM encoder for prompt compression, is calculated as~\cite{faiz2024llmcarbon}:
\begin{equation}
E_e = t_{\text{e}}^{{\rm SLM}} \left( \kappa \right) n_{\text{gpu}}^{{\rm SLM}} P_{\text{gpu}}^{{\rm SLM}} + t_{\text{e}}^{{\rm LLM}} \left( \kappa \right) n_{\text{gpu}}^{{\rm LLM}} P_{\text{gpu}}^{{\rm LLM}},
\end{equation}
where $t_{\text{e}}^{{\rm SLM}}$ represents the GPU execution time in SLM, $n_{\text{gpu}}$ denotes the number of GPUs utilized, and $P_{\text{gpu}}$ is the thermal design power per GPU, and superscript ${\rm LLM}$ denotes the corresponding parameters for the LLM.

The transmission energy consumption $E_t$ is given by:
\begin{equation}
E_t = t_{\text{t}} \left( \kappa \right) P_{\text{T}} = \frac{s\left( \kappa \right) }{R} P_{\text{T}},
\end{equation}
where $s$ represents the bit length of the compressed prompt $\hat{\textit{\textbf{x}}}$ with compression ratio $\kappa$, and $R$ is the transmission rate that can be expressed as:
\begin{equation}
R = W\log_2 \left(1 + \gamma \right) = W\log_2 \left(1 + \frac{P_{\text{T}}gd^{-\omega}}{\lambda^2} \right),
\end{equation}
where $W$ is the bandwidth of the offloading link between the user and DCO, $\gamma$ is the SNR, $g$ is the Rayleigh fading coefficient (exponentially distributed with unit mean), $d$ represents the distance between user and DCO, $\omega$ is the path-loss exponent, and $\lambda^2$ represents the Gaussian noise term in the Additive White Gaussian Noise (AWGN) channel.

\subsection{Service Delay and Error}
The total time consumption $T$ comprises three components: encoding delay in both SLM and LLM, and transmission delay, as is given by:
\begin{equation}
T \left( \kappa, P_T \right) = t_{\text{e}}^{{\rm SLM}} \left( \kappa \right) + t_{\text{e}}^{{\rm LLM}} \left( \kappa \right) + t_{\text{t}}\left( \kappa, P_T \right).
\end{equation}
Beyond maintaining the delay within acceptable bounds, we must also consider potential service degradation caused by transmission errors in the wireless channel. 

For wireless network-aided LLM services, user prompts must be uploaded to the LLM through a wireless environment. Due to potential bit errors during wireless transmission, these errors can directly affect the fidelity of the transmitted data. Therefore, it is crucial to consider the BEP to reflect the likelihood that the textural prompt will be incorrectly received or decoded in the wireless system.
For the $n_{\rm th}$ user, the average BEP $\eta$ can be given under various modulation formats by~\cite{tse2005fundamentals}:
\begin{equation}\label{BEREQUATION}
{\rm BEP} = \int_0^\infty  {\frac{{\zeta \ \left( {{\mu _2},{\mu _1}\tau } \right)}}{{2\zeta \ \left( {{\mu _2}} \right)}}{\varphi_{\tau_n}}\left( \tau  \right)d\tau },
\end{equation}
where $\zeta \left( { \cdot , \cdot } \right)$ is the upper incomplete Gamma function \cite[eq. (8.350.2)]{gradshteyn2007}, $\varphi$ represents the Probability Density Function (PDF) expression for wireless channel fading.  $ {{{\zeta \left( {{\mu _2},{\mu _1}\tau } \right)}}/{{2\zeta \left( {{\mu _2}} \right)}}} $ is the conditional bit-error probability, ${\mu _1} $ and ${\mu _2}$ are parameters specific to the modulation scheme, representing different combinations of modulation and detection techniques.


\section{Joint Power and Prompt Optimization}
This section introduces the JPPO++ for wireless network-aided LLM services. After formulating the problem, we propose a Double Deep Q-Network (DQN) method~\cite{van2016deep} to address the joint optimization problem. 

\subsection{Problem Formulation}
For our JPPO++ framework, we formulate an optimization problem that balances three key aspects: prompt compression quality, wireless transmission efficiency, and LLM service performance. The objective is to maximize the overall fidelity while satisfying power and latency constraints in the wireless network-aided LLM service system. Specifically, the joint optimization problem can be formulated as:
\begin{align}
\mathop
{\max}\limits_{\{ \kappa, P_T \}} &\
\mathbf{f}\left(\kappa, \eta\left(P_T\right)\right), \label{equ:consuser}\\
\text{s.t.}
&~E \left(\kappa, P_T\right) \le E_{\text{th}}, \tag{\ref{equ:consuser}{a}}\label{equ:consuser d}\\
&~P_{\text{T}} \le P_{\text{th}}, \tag{\ref{equ:consuser}{b}}\label{equ:consuser a}\\
&~T \le T_{\text{th}},\tag{\ref{equ:consuser}{c}}\label{equ:consuser b}\\
&~\mathbf{f} > \mathbf{f}_{\text{th}},\tag{\ref{equ:consuser}{d}}\label{equ:consuser c}
\end{align}
where $P_{\text{th}}$ is the maximum allowable power consumption, $T_{\text{th}}$ represents the maximum tolerable end-to-end latency, and $\mathbf{f}_{\text{th}}$ defines the minimum required fidelity. The constraints are designed to ensure the practical operation of the system. 
Constraint~\eqref{equ:consuser d} is the energy constraint that represents the total energy budget limitation at the edge device side, introducing a critical trade-off: while higher transmission power $P_T$ can lead to lower BEP $\eta$ and thus improved wireless transmission fidelity $\mathbf{f_2}$ to enhance the overall fidelity $\mathbf{f}$, the energy constraint forces a higher compression ratio $\kappa$ to reduce both the SLM and LLM inference energy cost and wireless transmission energy consumption. However, an excessively high compression ratio can result in significant information loss from the original prompt, potentially degrading both the semantic preservation fidelity $\mathbf{f_1}$ and LLM service quality fidelity $\mathbf{f_3}$.
Constraint~\eqref{equ:consuser a} ensures the power consumption remains within the device's power budget, Constraint~\eqref{equ:consuser b} guarantees that the total service latency meets real-time requirements, and Constraint~\eqref{equ:consuser c} maintains the quality of service by enforcing a minimum threshold on fidelity. This optimization framework allows us to find the optimal balance between compression ratio and transmission power while maintaining high-quality LLM service delivery.


\subsection{Double DQN Solution}
To solve the complex JPPO++ optimization problem, we deploy a centralized Double DQN method to find optimal prompt compression and transmission strategies for $N$ users, as shown in~\textbf{Algorithm 1}. 
\subsubsection{Algorithm Design}
The centralized control of the DDQN agent ensures fairness by dynamically allocating bandwidth and power based on real-time user demands and varying channel conditions. The key elements of the proposed Double DQN design are given as follows:
\begin{itemize}
\item \emph{Environment:} The environment of the Double DQN algorithm in the proposed framework is the communication environment with $N$ users.
\item \emph{State:} The state information includes the current fidelity of the transmitted message, SNR, and BEP. The state information of the $n_{\rm th}$ user is captured in a 3-dimensional vector: $\left[\mathbf{f}_n(\eta_n), \gamma_n, \right]$.
\item \emph{Action:} The actions include selecting compression and power levels. The action space is denoted as $\mathcal{A}=\{\mathcal{A}_1,\ldots, \mathcal{A}_N\}$, where $\mathcal{A}_n$ is the action of the $n_{\rm th}$ user and consists of a tuple with discrete values of compression ratio and transmission power level. The compression ratio level is a discrete value range from $0$ to $4$. The transmission power level is a discrete value ranging from $0$ to $9$, which affects BEP.

\item \emph{Reward:} The reward of the $n_{\rm th}$ user is $\mathcal{R}_n$. In each episode, the agent accumulates rewards based on its actions. The reward maximizes fidelity while minimizing penalties related to BEP and power usage.
\end{itemize}

\begin{algorithm}[t]
\caption{Double DQN Algorithm}
\label{alg:double_dqn}
\hspace*{0.02in} {\bf Input:}
\begin{itemize}
\item Action space $\mathcal{A} = \{\mathcal{A}_1,\ldots,\mathcal{A}_N\}$
\item Target privacy parameters $(\epsilon, \delta)$
\item Learning rate $\xi$, exploration decay $\epsilon_{\text{decay}}$
\end{itemize}
\hspace*{0.02in} {\bf Output:}
\begin{itemize}
\item Learned Q-network with optimized policy
\end{itemize}
\begin{algorithmic}[1]
\State Initialize Q-network and target Q-network with random weights
\State Initialize replay buffer and initial state $s_0$
\For{each episode $k \in \{1, \dots, K\}$}
\State Reset environment and observe initial state $s_0$
\While{not episode terminated}
\State Select action $a_t$ using $\epsilon$-greedy policy from Q-network
\State Execute $a_t$ and observe reward $r_t$, and then go to next state $s_{t+1}$
\State Store transition $(s_t, a_t, r_t, s_{t+1})$ into replay buffer
\State $s_t \leftarrow s_{t+1}$

\State \textit{\# Training step:}
\State Sample mini-batch of transitions $(s, a, r, s')$ from replay buffer
\For{each transition in the mini-batch}
\If{$s'$ is terminal}
\State $y \leftarrow r$
\Else
\State Go to~(\ref{update})
\EndIf
\State Compute loss $\mathcal{L} = \left(y - Q\left(s, a; \theta\right)\right)^2$
\State Perform gradient descent step on $\mathcal{L}$ to update Q-network
\EndFor

\State Decay exploration rate: $\epsilon \leftarrow \max\left(\epsilon \cdot \epsilon_{\text{decay}}, \epsilon_{\min}\right)$
\State Periodically update target Q-network: $\theta^- \leftarrow \theta$
\EndWhile
\EndFor
\end{algorithmic}
\end{algorithm}

The key design of the Double DQN is to decouple action selection from evaluation and address the over-estimation issue by using two separate networks: {\textit{Current Q-network}}, which predicts Q-values based on the current state, and {\textit{Target Q-network}}, which calculates target Q-values during updates and evaluates the Q-value of the best next action selected by the current Q-network, making the learning process of the Double DQN more stable.

The update steps of DQN are:  
\begin{equation} 
Q\left(s_t, a_t\right) \leftarrow Q\left(s_t, a_t\right) + \xi r_{t+1} + \mu \max_{a} Q(s_{t+1}, a) - Q\left(s_t, a_t\right),
\end{equation}  
where $ Q\left(s_t, a_t\right) $ is the estimated Q-value updated by the Bellman equation for taking action $a_t$ in state $s_t$. $ \xi $ is the learning rate. $ r_{t+1} $ is the reward received after taking action $ a_t $ in state $s_t$ and transitioning to state $ s_{t+1} $. $ \mu $ is the discount factor. 
And the loss function is the Mean Squared Error (MSE) between the predicted Q-values and the target Q-values, as is given by:  
\begin{equation} 
L_i(\theta_i) = \left[ r_{t+1} + \mu \max_{a'} Q_{\text{target}}(s_{t+1}, a'; \theta_i^*) - Q(s_t, a_t; \theta_i) \right]^2, 
\end{equation}  
where $ L_i(\theta_i) $ is the loss for the $ i $-th iteration with parameter $ \theta_i $. $ r_{t+1}$ and $ s_{t+1}$ are the reward and next state observed after taking action $ a_t $ in state $s_t$. $ Q_{\text{target}}(s_{t+1}, a'; \theta_i^*) $ is the target Q-value estimated by the target network with parameter $ \theta_i^* $.  $ Q(s_t, a_t; \theta_i) $ is the predicted Q-value by current Q-network with parameter $ \theta_i $. 


Then the update of the Double DQN can be given by:
\begin{equation}\label{update}
y = r + \mu Q_{\text{target}} \left( s', \arg\max_{a'} Q(s', a'; \theta); \theta^- \right),
\end{equation}
where $Q(s, a; \theta)$ is the estimated Q-value from the current Q-network with parameter $\theta$ and $Q_{\text{target}} ( s', a'; \theta^-)$ is the Q-value from the target network with parameter $\theta^-$.

\subsubsection{Complexity Analysis}
The computational complexity of the Double DQN algorithm is primarily determined by the number of training episodes, the size of the neural network, and the batch operations performed during training. Let $K$ denote the number of training episodes, $\mathcal{T}$ denote the average number of time steps per episode, $B$ denote the batch size, and $\mathcal{P}$ denote the number of parameters in the Q-network.

At each time step, the algorithm performs a forward pass through the Q-network for action selection, taking $\mathcal{O}(\mathcal{P})$ time. When updating the Q-network, a mini-batch of $B$ transitions is sampled from the replay buffer. For each sample, forward passes through both the current and target Q-networks, as well as a backward pass for optimization, are required. These operations contribute a per-step training cost of $\mathcal{O}(B \cdot \mathcal{P})$. Therefore, the total time complexity over all episodes is $\mathcal{O}(K \cdot \mathcal{T} \cdot B \cdot \mathcal{P})$.

In terms of space, the main contributors are the Q-network parameters, requiring $\mathcal{O}(\mathcal{P})$ memory, and the replay buffer, which stores up to $M$ transitions of size proportional to the state and action dimensions. Assuming each transition requires $\mathcal{O}(d)$ space, where $d$ is the dimension of the state vector, the total space complexity is $\mathcal{O}(\mathcal{P} + M \cdot d)$.

\section{Numerical Results}\label{nums}
We consider a practical simulation environment with variable fidelity, SNR, and BEP to support a wireless network-aided LLM service framework. 
The centralized DQN agent manages the environment, which involves selecting compression and power levels. Its goal is to balance fidelity, minimize errors, and optimize power usage. The trained DRL model is designed with a modular architecture, making it adaptable to other types of networks and tasks. Its state representation, action selection, and reward mechanisms can be independently modified to accommodate communication networks. 

\subsection{Experimental Setup}
Following the LLMLingua platform, we employ the SLM, i.e., {\textit{GPT-Neo 125M}}, for prompt compression and the LLM, i.e., {\textit{GPT-J 6B}}, to generate responses for users~\cite{rothman2022transformers}. The simulations are carried out based on the {\textit{MeetingBank-transcript}} dataset~\cite{hu-etal-2023-meetingbank} and {\textit{LongBench}} datset ~\cite{bai2023longbench}, where we select the long prompt data whose length exceeds $500$ tokens from the dataset to apply in our system simulations. Other parameter settings are in Table~\ref{tab:sim_params}.
\begin{table}[t]
\centering
\caption{Simulation Parameter Configuration}
\label{tab:sim_params}
{\small
\begin{tabular}{l|c}
\toprule
\textbf{Parameter} & \textbf{Value} \\
\midrule
Learning Rate & $10^{-3}$ \\
$\alpha_1$, $\alpha_2$, $\alpha_3$ & $0.4$, $0.3$, $0.3$ \\
Total Test Runs Range & $\left[1, 10\right]$ \\
Episodes per Test Run & $10{,}000$ \\
\bottomrule
\end{tabular}}
\end{table}

The details of the  datasets and experimental conditions are given as follows:
\begin{itemize}
\item \textbf{Network Conditions:} Our simulated MEC environment includes a total bandwidth of $20$ MHz, a maximum transmission power of $10$ dBm per user. In the MEC environment, the transmission power ranges from a minimum of 1 dBm (approximately 0.00126 W) to a maximum of 10 dBm (0.01 W). We  divide the range of power into 10 power levels evenly, where each level increases by 1 dBm, corresponding to a gradual increase in power from low to high. Additionally, the channel quality of the network varies dynamically in the range of $0.1$ to $1$.

\item \textbf{User Demand:} User task demands are randomly initialized, simulating heterogeneous data processing requirements.

\item \textbf{Step Limit:} Each episode is capped at $100$ steps to simulate practical usage constraints.

\item \textbf{Random Seed:} To ensure reproducibility, we have initialized random seeds for both NumPy and PyTorch, ensuring consistency in experimental results.
\end{itemize}

\begin{figure*}[t]
\centering
\includegraphics[width=0.9\textwidth]{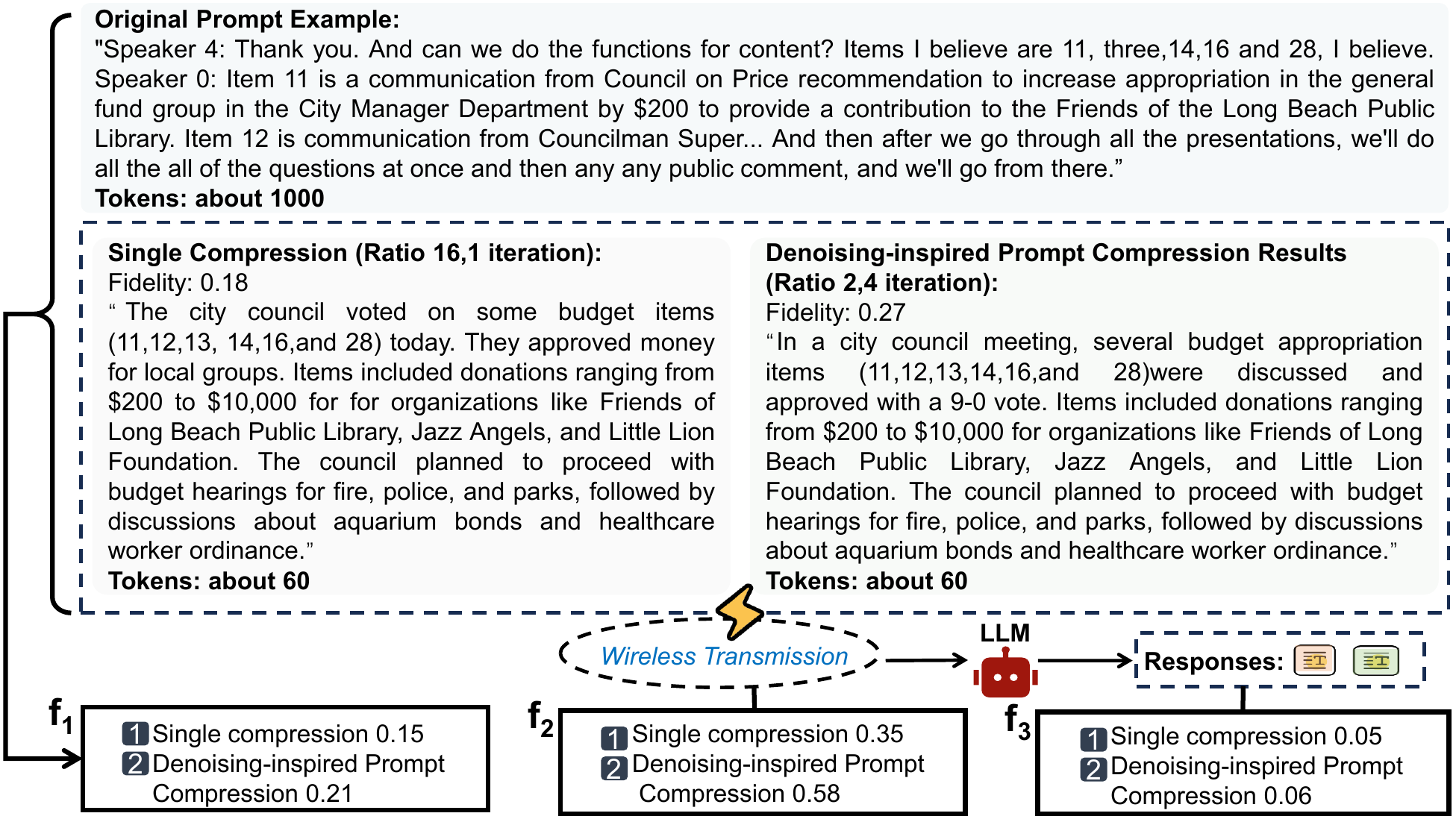}
\caption{The example illustrates wireless network-aided LLM services with SLM-based prompt compression, with a compression example of a single $16\times$ compression ratio and an iterative $4$ times $2\times$ compression ratio.}
\label{examplefa}
\end{figure*}

\subsection{Experiments Performance Analysis}
To demonstrate the effectiveness and generalizability of our proposed JPPO++ framework, we design a sequence of experiments to answer the following research questions:
\begin{itemize}
\item \textbf{Q1. Effectiveness:} Does denoising-inspired prompt compression improve fidelity compared to one-shot compression?
\item \textbf{Q2. Necessity:} How do compression ratio and transmission power jointly affect QoS in network-aided LLM services?
\item \textbf{Q3. Adaptivity:} Can DRL effectively optimize prompt compression and power control under dynamic network and task conditions?
\end{itemize}

\subsubsection{Q1. Effectiveness}
Fig.~\ref{examplefa} compares prompt compression using a single-round $16\times$ compression ratio with an iterative method applying $2\times$ compression over 4 times. Based on the fidelity metric $\mathbf{f}$, the iterative strategy consistently outperforms the single-step counterpart across all three components, demonstrating better semantic preservation and communication robustness. The iterative method achieves a final fidelity score of $0.27$, compared to $0.18$ for the single-round compression. The total processing time is $5.90$ seconds, including $2.92$ seconds for compression, $2.48$ seconds for LLM response generation, and $0.50$ seconds for transmission. From the DRL perspective, it yields a higher reward score of $26.40$ while consuming $8.18$ units of power.

In contrast, the single-round $16\times$ compression approach consumes less power (i.e., $2.90$ units), but produces a lower reward of $17.74$ and a longer total processing time of $11.25$ seconds that is nearly double that of the iterative method. This includes $6.59$ seconds for compression, $4.12$ seconds for response generation, and $0.55$ seconds for transmission. These results confirm that denoising-inspired iterative compression in JPPO++ not only improves fidelity but also significantly reduces end-to-end latency, making it a more efficient strategy for LLM prompt compression under wireless constraints.
We further validate the effectiveness of our framework using wider tests in the {\textit{MeetingBank-transcript}} dataset~\cite{hu-etal-2023-meetingbank}, where prompts exceed $500$ tokens. When prioritizing aggressive compression while tolerating up to a $30\%$ fidelity drop, JPPO++ supports compression ratios up to $16\times$. Under this setting, service time is significantly reduced. The single-round baseline achieves a $42.3\%$ reduction in total service time compared to the no-compression case, while our iterative denoising-based method improves it further to $46.5\%$.

\subsubsection{Q2. Necessity}
\begin{figure}[ht]
\centering
\setlength{\tabcolsep}{2.5pt}
\scriptsize
\begin{minipage}{0.45\textwidth}
\centering
\captionof{table}{One-shot reward values of the baseline single-round compression method.}
\vspace{-0.2cm}
\label{tab:0}
\begin{tabular}{cc|cccccccccc}
\hline
& & \multicolumn{10}{c}{\textbf{Power Level}} \\
& & \textbf{1} & \textbf{2} & \textbf{3} & \textbf{4} & \textbf{5} & \textbf{6} & \textbf{7} & \textbf{8} & \textbf{9} & \textbf{10} \\
\hline
\multirow{10}{*}{\rotatebox{90}{\textbf{Compression}}} & \textbf{1} & \cellcolor[RGB]{243,243,248}10.11 & \cellcolor[RGB]{229,228,240}10.64 & \cellcolor[RGB]{225,224,238}10.76 & \cellcolor[RGB]{227,226,239}10.71 & \cellcolor[RGB]{230,230,241}10.58 & \cellcolor[RGB]{235,236,245}10.39 & \cellcolor[RGB]{241,241,247}10.18 & \cellcolor[RGB]{246,246,250}9.95 & \cellcolor[RGB]{251,250,252}9.71 & \cellcolor[RGB]{254,253,254}9.45 \\
& \textbf{2} & \cellcolor[RGB]{215,214,233}11.12 & \cellcolor[RGB]{205,193,222}11.66 & \cellcolor[RGB]{202,188,220}11.77 & \cellcolor[RGB]{203,190,221}11.72 & \cellcolor[RGB]{206,195,224}11.59 & \cellcolor[RGB]{210,202,227}11.41 & \cellcolor[RGB]{214,211,231}11.20 & \cellcolor[RGB]{220,219,235}10.97 & \cellcolor[RGB]{226,225,239}10.72 & \cellcolor[RGB]{233,234,243}10.47 \\
& \textbf{3} & \cellcolor[RGB]{210,202,227}11.40 & \cellcolor[RGB]{199,182,218}11.93 & \cellcolor[RGB]{197,178,215}12.05 & \cellcolor[RGB]{198,180,216}12.00 & \cellcolor[RGB]{200,185,219}11.87 & \cellcolor[RGB]{204,191,222}11.69 & \cellcolor[RGB]{208,200,226}11.47 & \cellcolor[RGB]{213,209,230}11.24 & \cellcolor[RGB]{219,218,235}11.00 & \cellcolor[RGB]{226,225,238}10.75 \\
& \textbf{4} & \cellcolor[RGB]{210,202,227}11.41 & \cellcolor[RGB]{199,182,217}11.94 & \cellcolor[RGB]{201,146,241}12.06 & \cellcolor[RGB]{198,179,216}12.01 & \cellcolor[RGB]{200,185,219}11.87 & \cellcolor[RGB]{204,191,221}11.69 & \cellcolor[RGB]{208,199,225}11.48 & \cellcolor[RGB]{213,209,230}11.25 & \cellcolor[RGB]{218,218,235}11.01 & \cellcolor[RGB]{226,224,238}10.75 \\
& \textbf{5} & \cellcolor[RGB]{212,206,229}11.31 & \cellcolor[RGB]{201,185,219}11.85 & \cellcolor[RGB]{199,181,217}11.96 & \cellcolor[RGB]{199,183,218}11.91 & \cellcolor[RGB]{202,188,220}11.78 & \cellcolor[RGB]{206,194,223}11.60 & \cellcolor[RGB]{210,203,227}11.39 & \cellcolor[RGB]{215,212,232}11.16 & \cellcolor[RGB]{221,220,236}10.91 & \cellcolor[RGB]{228,227,240}10.66 \\
& \textbf{6} & \cellcolor[RGB]{214,211,231}11.19 & \cellcolor[RGB]{203,190,221}11.72 & \cellcolor[RGB]{201,186,219}11.84 & \cellcolor[RGB]{202,188,220}11.79 & \cellcolor[RGB]{205,193,222}11.65 & \cellcolor[RGB]{208,200,226}11.47 & \cellcolor[RGB]{213,208,230}11.26 & \cellcolor[RGB]{218,217,234}11.03 & \cellcolor[RGB]{225,224,238}10.79 & \cellcolor[RGB]{232,232,242}10.53 \\
& \textbf{7} & \cellcolor[RGB]{217,216,234}11.04 & \cellcolor[RGB]{206,196,224}11.58 & \cellcolor[RGB]{204,191,221}11.69 & \cellcolor[RGB]{205,193,222}11.64 & \cellcolor[RGB]{208,198,225}11.51 & \cellcolor[RGB]{211,205,228}11.33 & \cellcolor[RGB]{216,214,233}11.12 & \cellcolor[RGB]{222,221,236}10.89 & \cellcolor[RGB]{228,228,240}10.64 & \cellcolor[RGB]{236,236,245}10.39 \\
& \textbf{8} & \cellcolor[RGB]{222,221,236}10.90 & \cellcolor[RGB]{209,201,226}11.43 & \cellcolor[RGB]{207,197,224}11.55 & \cellcolor[RGB]{208,199,225}11.50 & \cellcolor[RGB]{211,204,228}11.36 & \cellcolor[RGB]{214,212,232}11.18 & \cellcolor[RGB]{219,218,235}10.97 & \cellcolor[RGB]{226,225,238}10.74 & \cellcolor[RGB]{233,233,243}10.50 & \cellcolor[RGB]{240,240,247}10.24 \\
& \textbf{9} & \cellcolor[RGB]{226,224,238}10.75 & \cellcolor[RGB]{212,207,230}11.28 & \cellcolor[RGB]{210,202,227}11.40 & \cellcolor[RGB]{211,204,228}11.35 & \cellcolor[RGB]{214,210,231}11.21 & \cellcolor[RGB]{218,217,234}11.03 & \cellcolor[RGB]{224,222,237}10.82 & \cellcolor[RGB]{230,230,241}10.59 & \cellcolor[RGB]{237,237,245}10.35 & \cellcolor[RGB]{244,244,249}10.09 \\
& \textbf{10} & \cellcolor[RGB]{230,229,241}10.60 & \cellcolor[RGB]{215,214,233}11.13 & \cellcolor[RGB]{213,209,230}11.25 & \cellcolor[RGB]{214,211,231}11.20 & \cellcolor[RGB]{217,216,234}11.07 & \cellcolor[RGB]{222,221,236}10.88 & \cellcolor[RGB]{228,227,240}10.67 & \cellcolor[RGB]{234,234,244}10.44 & \cellcolor[RGB]{241,241,247}10.20 & \cellcolor[RGB]{247,246,250}9.94 \\
\hline
\end{tabular}
\end{minipage}
\vspace{0.5cm}

\begin{minipage}{0.45\textwidth}
\centering
\captionof{table}{One-shot reward values of the denoising-inspired prompt compression method with linear schedule.}
\label{tab:1}
\begin{tabular}{cc|cccccccccc}
\hline
& & \multicolumn{10}{c}{\textbf{Power Level}} \\
& & \textbf{1} & \textbf{2} & \textbf{3} & \textbf{4} & \textbf{5} & \textbf{6} & \textbf{7} & \textbf{8} & \textbf{9} & \textbf{10} \\
\hline
\multirow{10}{*}{\rotatebox{90}{\textbf{Compression}}} & \textbf{1} & \cellcolor[RGB]{254,253,254}10.25 & \cellcolor[RGB]{240,240,247}10.93 & \cellcolor[RGB]{232,232,243}11.17 & \cellcolor[RGB]{230,229,241}11.25 & \cellcolor[RGB]{230,230,241}11.24 & \cellcolor[RGB]{232,232,243}11.17 & \cellcolor[RGB]{236,236,245}11.07 & \cellcolor[RGB]{240,240,247}10.94 & \cellcolor[RGB]{244,244,249}10.79 & \cellcolor[RGB]{248,247,251}10.63 \\
& \textbf{2} & \cellcolor[RGB]{229,229,241}11.27 & \cellcolor[RGB]{210,203,227}11.94 & \cellcolor[RGB]{205,193,222}12.18 & \cellcolor[RGB]{203,189,221}12.26 & \cellcolor[RGB]{203,190,221}12.24 & \cellcolor[RGB]{205,193,222}12.18 & \cellcolor[RGB]{207,197,225}12.07 & \cellcolor[RGB]{210,203,227}11.94 & \cellcolor[RGB]{214,210,231}11.79 & \cellcolor[RGB]{218,217,235}11.63 \\
& \textbf{3} & \cellcolor[RGB]{221,220,236}11.54 & \cellcolor[RGB]{204,191,222}12.21 & \cellcolor[RGB]{198,181,217}12.46 & \cellcolor[RGB]{197,178,215}12.53 & \cellcolor[RGB]{197,178,216}12.51 & \cellcolor[RGB]{199,181,217}12.44 & \cellcolor[RGB]{201,186,219}12.34 & \cellcolor[RGB]{204,191,222}12.21 & \cellcolor[RGB]{208,198,225}12.06 & \cellcolor[RGB]{212,206,229}11.89 \\
& \textbf{4} & \cellcolor[RGB]{220,219,236}11.55 & \cellcolor[RGB]{204,191,221}12.21 & \cellcolor[RGB]{198,181,217}12.46 & \cellcolor[RGB]{201,146,241}12.53 & \cellcolor[RGB]{197,178,216}12.51 & \cellcolor[RGB]{199,181,217}12.44 & \cellcolor[RGB]{201,186,219}12.34 & \cellcolor[RGB]{204,191,222}12.20 & \cellcolor[RGB]{208,198,225}12.05 & \cellcolor[RGB]{212,206,229}11.88 \\
& \textbf{5} & \cellcolor[RGB]{224,222,237}11.45 & \cellcolor[RGB]{206,195,224}12.12 & \cellcolor[RGB]{200,185,219}12.36 & \cellcolor[RGB]{199,182,217}12.43 & \cellcolor[RGB]{199,183,218}12.41 & \cellcolor[RGB]{201,186,219}12.34 & \cellcolor[RGB]{203,190,221}12.23 & \cellcolor[RGB]{207,196,224}12.10 & \cellcolor[RGB]{210,203,227}11.95 & \cellcolor[RGB]{214,211,231}11.78 \\
& \textbf{6} & \cellcolor[RGB]{228,227,239}11.32 & \cellcolor[RGB]{209,201,226}11.99 & \cellcolor[RGB]{203,190,221}12.23 & \cellcolor[RGB]{202,188,220}12.30 & \cellcolor[RGB]{202,188,220}12.28 & \cellcolor[RGB]{204,191,222}12.20 & \cellcolor[RGB]{207,196,224}12.09 & \cellcolor[RGB]{210,202,227}11.96 & \cellcolor[RGB]{213,209,231}11.80 & \cellcolor[RGB]{218,217,235}11.63 \\
& \textbf{7} & \cellcolor[RGB]{232,232,242}11.18 & \cellcolor[RGB]{213,208,230}11.84 & \cellcolor[RGB]{207,197,224}12.08 & \cellcolor[RGB]{205,194,223}12.15 & \cellcolor[RGB]{206,195,223}12.13 & \cellcolor[RGB]{208,198,225}12.05 & \cellcolor[RGB]{210,203,227}11.94 & \cellcolor[RGB]{213,209,231}11.80 & \cellcolor[RGB]{217,216,234}11.65 & \cellcolor[RGB]{223,222,237}11.47 \\
& \textbf{8} & \cellcolor[RGB]{237,237,245}11.03 & \cellcolor[RGB]{216,215,233}11.69 & \cellcolor[RGB]{210,204,228}11.93 & \cellcolor[RGB]{209,201,226}12.00 & \cellcolor[RGB]{209,202,227}11.97 & \cellcolor[RGB]{211,205,228}11.90 & \cellcolor[RGB]{214,210,231}11.78 & \cellcolor[RGB]{217,216,234}11.65 & \cellcolor[RGB]{222,221,237}11.49 & \cellcolor[RGB]{228,227,240}11.31 \\
& \textbf{9} & \cellcolor[RGB]{242,242,248}10.88 & \cellcolor[RGB]{221,220,236}11.54 & \cellcolor[RGB]{214,211,231}11.78 & \cellcolor[RGB]{213,208,230}11.84 & \cellcolor[RGB]{213,209,230}11.82 & \cellcolor[RGB]{215,213,232}11.74 & \cellcolor[RGB]{218,217,235}11.62 & \cellcolor[RGB]{222,221,237}11.48 & \cellcolor[RGB]{228,227,239}11.32 & \cellcolor[RGB]{233,233,243}11.15 \\
& \textbf{10} & \cellcolor[RGB]{246,245,250}10.73 & \cellcolor[RGB]{226,224,238}11.39 & \cellcolor[RGB]{218,217,235}11.62 & \cellcolor[RGB]{216,215,233}11.69 & \cellcolor[RGB]{217,216,234}11.66 & \cellcolor[RGB]{219,218,235}11.58 & \cellcolor[RGB]{223,222,237}11.46 & \cellcolor[RGB]{228,227,239}11.32 & \cellcolor[RGB]{233,233,243}11.16 & \cellcolor[RGB]{238,238,246}10.98 \\
\hline
\end{tabular}
\end{minipage}
\vspace{0.5cm}

\begin{minipage}{0.45\textwidth}
\centering
\captionof{table}{One-shot reward values of the denoising-inspired prompt compression method with cosine schedule.}
\label{tab:2}
\begin{tabular}{cc|cccccccccc}
\hline
& & \multicolumn{10}{c}{\textbf{Power Level}} \\
& & \textbf{1} & \textbf{2} & \textbf{3} & \textbf{4} & \textbf{5} & \textbf{6} & \textbf{7} & \textbf{8} & \textbf{9} & \textbf{10} \\
\hline
\multirow{10}{*}{\rotatebox{90}{\textbf{Compression}}} & \textbf{1} & \cellcolor[RGB]{233,234,243}9.96 & \cellcolor[RGB]{219,218,235}10.50 & \cellcolor[RGB]{216,215,234}10.62 & \cellcolor[RGB]{217,216,234}10.57 & \cellcolor[RGB]{221,220,236}10.44 & \cellcolor[RGB]{226,224,238}10.27 & \cellcolor[RGB]{231,231,242}10.06 & \cellcolor[RGB]{237,237,245}9.83 & \cellcolor[RGB]{244,244,249}9.59 & \cellcolor[RGB]{248,248,251}9.34 \\
& \textbf{2} & \cellcolor[RGB]{212,207,229}10.84 & \cellcolor[RGB]{202,187,220}11.38 & \cellcolor[RGB]{199,183,218}11.50 & \cellcolor[RGB]{200,184,218}11.45 & \cellcolor[RGB]{203,189,220}11.33 & \cellcolor[RGB]{206,195,223}11.15 & \cellcolor[RGB]{210,203,227}10.95 & \cellcolor[RGB]{214,212,232}10.72 & \cellcolor[RGB]{220,219,235}10.48 & \cellcolor[RGB]{226,225,239}10.24 \\
& \textbf{3} & \cellcolor[RGB]{209,202,227}10.98 & \cellcolor[RGB]{199,182,217}11.52 & \cellcolor[RGB]{201,146,241}11.64 & \cellcolor[RGB]{197,179,216}11.60 & \cellcolor[RGB]{200,184,218}11.48 & \cellcolor[RGB]{203,189,221}11.31 & \cellcolor[RGB]{207,197,224}11.10 & \cellcolor[RGB]{211,205,228}10.88 & \cellcolor[RGB]{216,214,233}10.65 & \cellcolor[RGB]{222,221,236}10.40 \\
& \textbf{4} & \cellcolor[RGB]{212,207,229}10.85 & \cellcolor[RGB]{201,186,219}11.39 & \cellcolor[RGB]{199,182,217}11.52 & \cellcolor[RGB]{200,183,218}11.48 & \cellcolor[RGB]{202,188,220}11.36 & \cellcolor[RGB]{205,194,223}11.19 & \cellcolor[RGB]{209,201,226}10.99 & \cellcolor[RGB]{213,209,231}10.78 & \cellcolor[RGB]{218,217,235}10.54 & \cellcolor[RGB]{225,224,238}10.30 \\
& \textbf{5} & \cellcolor[RGB]{216,215,234}10.62 & \cellcolor[RGB]{206,194,223}11.17 & \cellcolor[RGB]{203,189,221}11.30 & \cellcolor[RGB]{204,191,221}11.27 & \cellcolor[RGB]{206,195,224}11.15 & \cellcolor[RGB]{209,201,226}10.98 & \cellcolor[RGB]{213,209,230}10.79 & \cellcolor[RGB]{218,217,234}10.57 & \cellcolor[RGB]{224,222,237}10.34 & \cellcolor[RGB]{230,229,241}10.10 \\
& \textbf{6} & \cellcolor[RGB]{223,222,237}10.36 & \cellcolor[RGB]{211,204,228}10.92 & \cellcolor[RGB]{208,199,225}11.05 & \cellcolor[RGB]{209,200,226}11.02 & \cellcolor[RGB]{211,204,228}10.90 & \cellcolor[RGB]{214,211,231}10.74 & \cellcolor[RGB]{218,217,235}10.55 & \cellcolor[RGB]{224,223,237}10.33 & \cellcolor[RGB]{230,229,241}10.11 & \cellcolor[RGB]{236,236,245}9.87 \\
& \textbf{7} & \cellcolor[RGB]{230,230,241}10.10 & \cellcolor[RGB]{216,214,233}10.65 & \cellcolor[RGB]{213,209,230}10.79 & \cellcolor[RGB]{214,210,231}10.76 & \cellcolor[RGB]{216,214,233}10.65 & \cellcolor[RGB]{220,219,235}10.49 & \cellcolor[RGB]{225,224,238}10.30 & \cellcolor[RGB]{230,230,241}10.09 & \cellcolor[RGB]{236,237,245}9.86 & \cellcolor[RGB]{242,243,248}9.63 \\
& \textbf{8} & \cellcolor[RGB]{237,237,245}9.83 & \cellcolor[RGB]{222,221,237}10.38 & \cellcolor[RGB]{219,218,235}10.52 & \cellcolor[RGB]{220,219,235}10.50 & \cellcolor[RGB]{222,221,237}10.39 & \cellcolor[RGB]{226,225,239}10.23 & \cellcolor[RGB]{231,231,242}10.04 & \cellcolor[RGB]{237,237,245}9.84 & \cellcolor[RGB]{243,243,248}9.62 & \cellcolor[RGB]{247,247,250}9.39 \\
& \textbf{9} & \cellcolor[RGB]{244,244,249}9.56 & \cellcolor[RGB]{229,229,241}10.12 & \cellcolor[RGB]{226,225,238}10.26 & \cellcolor[RGB]{226,225,239}10.24 & \cellcolor[RGB]{229,229,241}10.13 & \cellcolor[RGB]{233,233,243}9.98 & \cellcolor[RGB]{238,238,246}9.79 & \cellcolor[RGB]{243,243,248}9.59 & \cellcolor[RGB]{248,247,250}9.37 & \cellcolor[RGB]{251,250,252}9.14 \\
& \textbf{10} & \cellcolor[RGB]{249,249,251}9.29 & \cellcolor[RGB]{237,237,245}9.85 & \cellcolor[RGB]{233,233,243}10.00 & \cellcolor[RGB]{233,233,243}9.98 & \cellcolor[RGB]{236,236,245}9.88 & \cellcolor[RGB]{240,240,247}9.72 & \cellcolor[RGB]{244,244,249}9.54 & \cellcolor[RGB]{248,247,251}9.34 & \cellcolor[RGB]{251,251,252}9.13 & \cellcolor[RGB]{254,253,254}8.90 \\
\hline
\end{tabular}
\end{minipage}
\vspace{0.5cm}

\begin{minipage}{0.45\textwidth}
\centering
\captionof{table}{One-shot reward values of the denoising-inspired prompt compression method with quadratic schedule.}
\label{tab:3}
\begin{tabular}{cc|cccccccccc}
\hline
& & \multicolumn{10}{c}{\textbf{Power Level}} \\
& & \textbf{1} & \textbf{2} & \textbf{3} & \textbf{4} & \textbf{5} & \textbf{6} & \textbf{7} & \textbf{8} & \textbf{9} & \textbf{10} \\
\hline
\multirow{10}{*}{\rotatebox{90}{\textbf{Compression}}} & \textbf{1} & \cellcolor[RGB]{254,253,254}4.44 & \cellcolor[RGB]{228,227,240}8.61 & \cellcolor[RGB]{215,213,232}10.29 & \cellcolor[RGB]{212,207,229}10.80 & \cellcolor[RGB]{213,208,230}10.68 & \cellcolor[RGB]{215,214,233}10.22 & \cellcolor[RGB]{220,219,236}9.56 & \cellcolor[RGB]{227,226,239}8.80 & \cellcolor[RGB]{233,233,243}7.98 & \cellcolor[RGB]{240,240,247}7.15 \\
& \textbf{2} & \cellcolor[RGB]{249,249,251}5.62 & \cellcolor[RGB]{218,217,234}9.92 & \cellcolor[RGB]{206,196,224}11.77 & \cellcolor[RGB]{202,188,220}12.45 & \cellcolor[RGB]{202,188,220}12.51 & \cellcolor[RGB]{204,191,221}12.22 & \cellcolor[RGB]{207,196,224}11.72 & \cellcolor[RGB]{210,203,227}11.10 & \cellcolor[RGB]{214,212,232}10.43 & \cellcolor[RGB]{219,218,235}9.72 \\
& \textbf{3} & \cellcolor[RGB]{247,247,250}6.01 & \cellcolor[RGB]{215,212,232}10.36 & \cellcolor[RGB]{203,190,221}12.27 & \cellcolor[RGB]{199,182,217}13.04 & \cellcolor[RGB]{198,181,217}13.18 & \cellcolor[RGB]{199,183,218}12.97 & \cellcolor[RGB]{202,187,220}12.55 & \cellcolor[RGB]{205,193,223}12.01 & \cellcolor[RGB]{209,200,226}11.40 & \cellcolor[RGB]{212,207,230}10.76 \\
& \textbf{4} & \cellcolor[RGB]{247,246,250}6.10 & \cellcolor[RGB]{214,211,231}10.46 & \cellcolor[RGB]{203,189,220}12.41 & \cellcolor[RGB]{198,180,217}13.22 & \cellcolor[RGB]{197,178,216}13.41 & \cellcolor[RGB]{198,180,216}13.25 & \cellcolor[RGB]{200,184,218}12.88 & \cellcolor[RGB]{203,189,221}12.38 & \cellcolor[RGB]{206,195,224}11.82 & \cellcolor[RGB]{210,202,227}11.21 \\
& \textbf{5} & \cellcolor[RGB]{247,247,250}6.09 & \cellcolor[RGB]{214,211,231}10.44 & \cellcolor[RGB]{203,189,220}12.40 & \cellcolor[RGB]{198,180,217}13.24 & \cellcolor[RGB]{201,146,241}13.46 & \cellcolor[RGB]{197,179,216}13.33 & \cellcolor[RGB]{199,183,218}12.99 & \cellcolor[RGB]{202,188,220}12.52 & \cellcolor[RGB]{205,193,223}11.99 & \cellcolor[RGB]{209,200,226}11.41 \\
& \textbf{6} & \cellcolor[RGB]{247,247,250}6.03 & \cellcolor[RGB]{215,212,232}10.36 & \cellcolor[RGB]{203,189,221}12.33 & \cellcolor[RGB]{198,181,217}13.18 & \cellcolor[RGB]{197,178,216}13.42 & \cellcolor[RGB]{198,179,216}13.31 & \cellcolor[RGB]{199,183,218}12.99 & \cellcolor[RGB]{202,187,220}12.55 & \cellcolor[RGB]{205,193,222}12.03 & \cellcolor[RGB]{208,199,225}11.48 \\
& \textbf{7} & \cellcolor[RGB]{248,247,250}5.96 & \cellcolor[RGB]{215,213,233}10.26 & \cellcolor[RGB]{204,191,221}12.23 & \cellcolor[RGB]{199,182,217}13.08 & \cellcolor[RGB]{197,179,216}13.34 & \cellcolor[RGB]{198,180,217}13.24 & \cellcolor[RGB]{200,183,218}12.94 & \cellcolor[RGB]{202,188,220}12.52 & \cellcolor[RGB]{205,193,222}12.02 & \cellcolor[RGB]{208,199,225}11.48 \\
& \textbf{8} & \cellcolor[RGB]{248,248,251}5.87 & \cellcolor[RGB]{216,214,233}10.15 & \cellcolor[RGB]{204,192,222}12.11 & \cellcolor[RGB]{199,183,218}12.97 & \cellcolor[RGB]{198,180,217}13.23 & \cellcolor[RGB]{198,181,217}13.15 & \cellcolor[RGB]{200,184,218}12.86 & \cellcolor[RGB]{202,188,220}12.45 & \cellcolor[RGB]{205,194,223}11.96 & \cellcolor[RGB]{208,200,226}11.43 \\
& \textbf{9} & \cellcolor[RGB]{248,248,251}5.79 & \cellcolor[RGB]{217,216,234}10.03 & \cellcolor[RGB]{205,193,223}11.98 & \cellcolor[RGB]{200,184,218}12.84 & \cellcolor[RGB]{199,181,217}13.11 & \cellcolor[RGB]{199,182,217}13.03 & \cellcolor[RGB]{201,185,219}12.75 & \cellcolor[RGB]{203,189,221}12.35 & \cellcolor[RGB]{206,194,223}11.88 & \cellcolor[RGB]{209,200,226}11.36 \\
& \textbf{10} & \cellcolor[RGB]{249,249,251}5.70 & \cellcolor[RGB]{218,217,234}9.91 & \cellcolor[RGB]{206,195,223}11.85 & \cellcolor[RGB]{201,186,219}12.71 & \cellcolor[RGB]{199,183,218}12.98 & \cellcolor[RGB]{200,184,218}12.91 & \cellcolor[RGB]{201,186,219}12.64 & \cellcolor[RGB]{204,191,221}12.24 & \cellcolor[RGB]{206,196,224}11.77 & \cellcolor[RGB]{209,202,227}11.27 \\
\hline
\end{tabular}
\end{minipage}
\end{figure}
Tables~\ref{tab:0}-\ref{tab:3} show the end-to-end DRL reward values of wireless network-aided LLM service under different combinations of compression levels and transmission power levels. Specifically, Table~\ref{tab:0} presents the results for baseline single-round $16\times$ compression, while Tables~\ref{tab:1}-\ref{tab:3} show the outcomes of denoising-inspired iterative compression with linear, cosine, and quadratic scheduling strategies, respectively. For each approach, the reward matrix reflects the impact of varying compression ratios (i.e., rows) and transmission power levels (i.e., columns) on final system performance.

We observe that, across all methods, the highest reward values are typically achieved at moderate compression levels (e.g., $2\times$-$5\times$) and moderate power levels (e.g., 2-5). This pattern reflects a fundamental trade-off shaped by energy constraints and semantic preservation. On the one hand, aggressive compression reduces prompt length and transmission overhead, enabling higher transmit power within the same energy budget. However, it also increases the risk of discarding critical semantic content, which in turn degrades the quality of LLM inference. On the other hand, lower compression ratios preserve more of the original prompt, thereby improving fidelity, but result in longer input sequences. This increases LLM inference latency and limits the available transmission power, potentially raising the BEP and reducing the overall reward. Similarly, power levels exhibit a non-linear effect: very low power leads to high BEP and unreliable transmission, while very high power may violate energy constraints. The observed reward surface, therefore, captures the complex interplay between compression aggressiveness and transmission reliability, highlighting the necessity of joint optimization.

Therefore, a key insight from this analysis is that compression ratio and power control are not independently tunable. Thus, joint optimization is necessary to strike a context-aware balance. Another critical observation lies in the differences among scheduling strategies. The denoising-inspired schemes with dynamic compression progression (particularly the quadratic and cosine schedules) achieve higher peak rewards compared to the single-shot baseline. For instance, the best-performing configuration under the quadratic schedule exceeds the baseline reward by over $10\%$. This indicates that gradually compressing prompts in stages, especially with curvature-adaptive schedules, better preserves key information and aligns more effectively with LLM inference patterns. These findings suggest that selecting a compression trajectory tailored to the prompt structure, according to real scenarios, can provide substantial performance benefits.

\subsubsection{Q3. Adaptivity}
\begin{figure}[t]
\centering
{\includegraphics[width=0.46\textwidth]{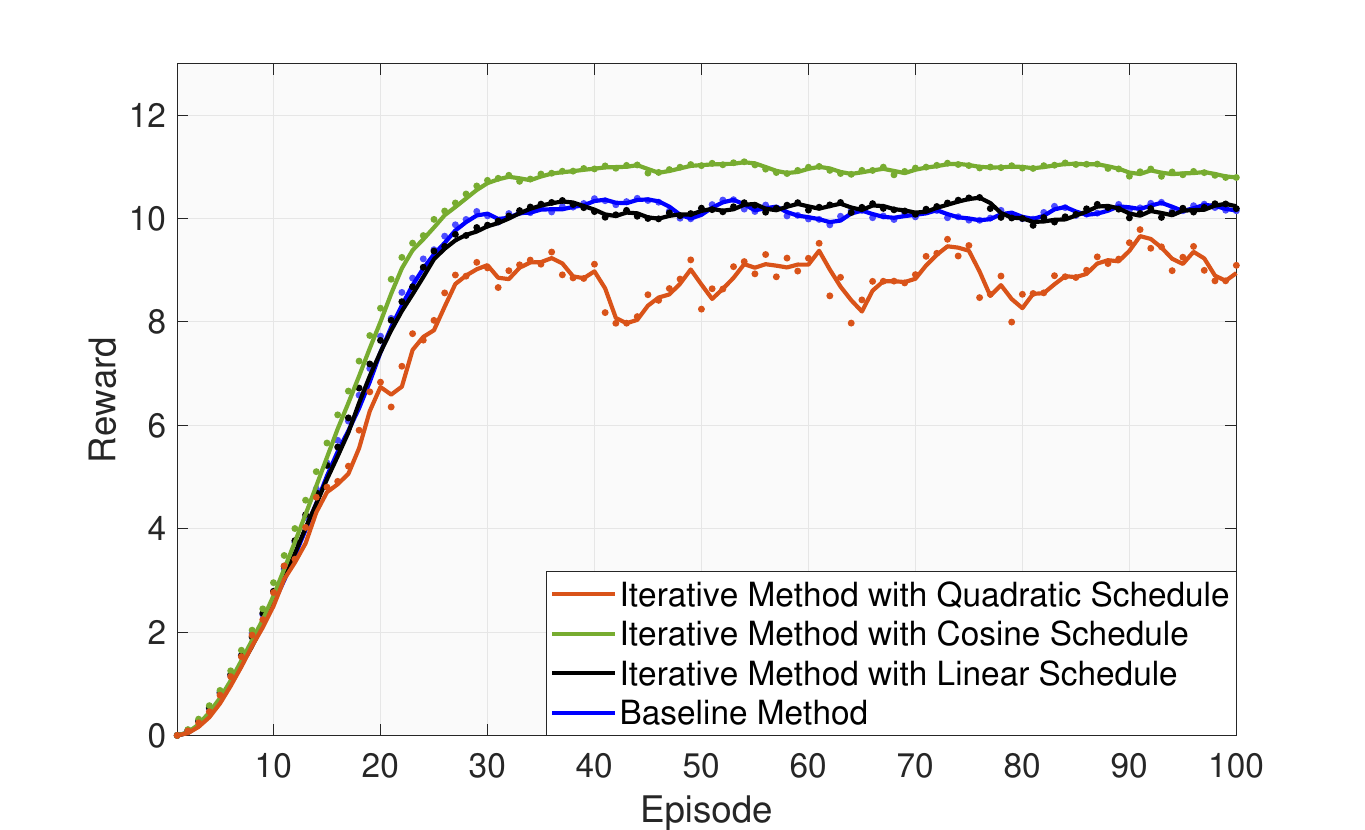}
\caption{The convergence performance of reward over 100 iterations for the DRL algorithm.}
\label{fig:result1}}
\end{figure}

Fig.~\ref{fig:result1} presents the reward convergence curves during the training process over $100$ training episodes using our Double DQN-based solution for the JPPO++ framework on long prompts from the {\textit{MeetingBank-transcript}} dataset~\cite{hu-etal-2023-meetingbank}. The DRL agent successfully learns optimal compression and power control policies under different iterative schedules. Notably, the cosine schedule consistently achieves the highest reward, demonstrating its effectiveness in preserving semantic coherence in long, structured dialogues. Its smooth transition function $\sigma(t) = \frac{1 - \cos(\pi t)}{2}$ enables gradual compression, with moderate intensity at the beginning and end and stronger compression in the middle. This pattern is particularly suited for meeting transcripts, where key information is interleaved with contextual background. Compared to the static baseline, the learned policy yields faster convergence and higher final performance, confirming that JPPO++ can adaptively balance fidelity and efficiency in real-time through policy learning.

\begin{figure}[t]
\centering
\subfigure[Transmission time reduction percentage]
{	\includegraphics[width=0.4\textwidth]{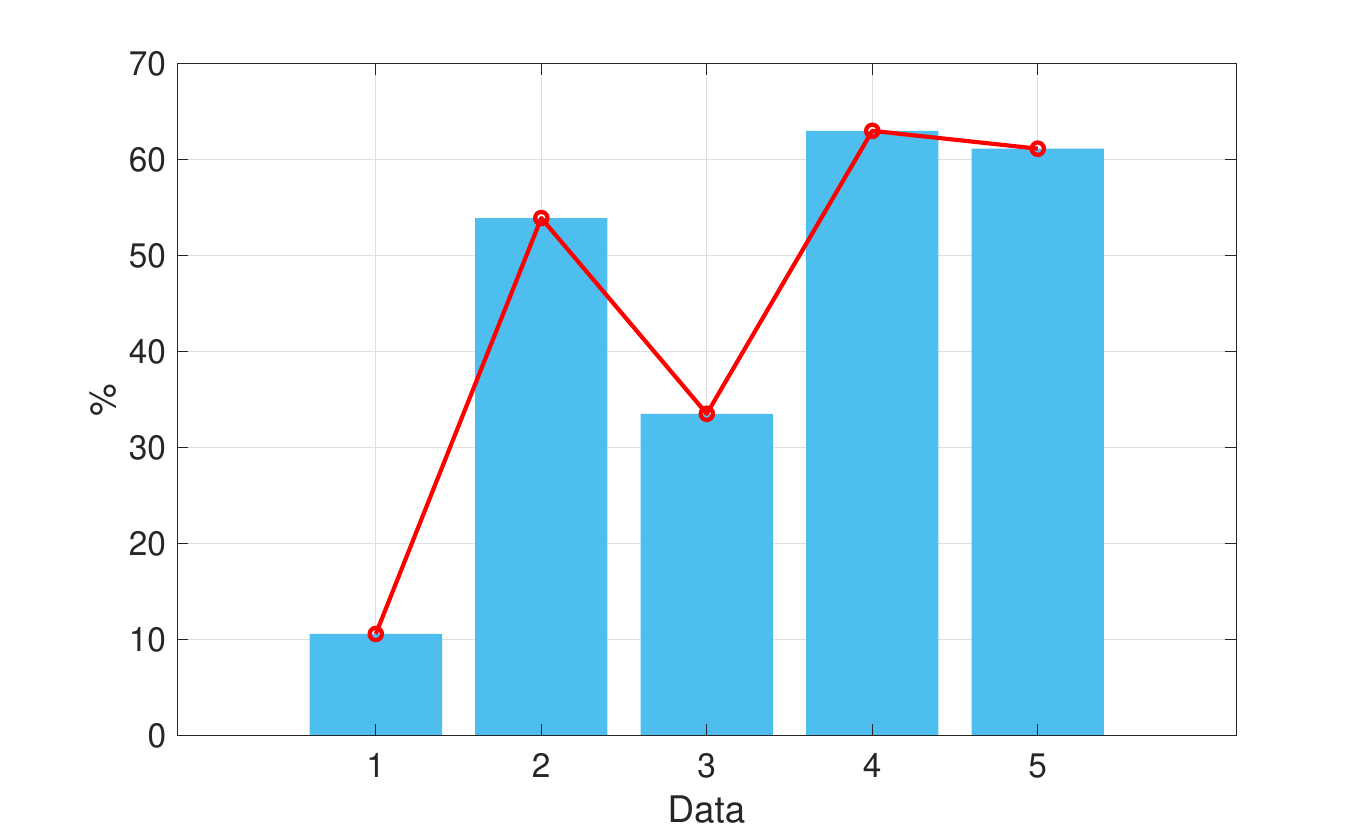}
}
\subfigure[Service time reduction percentage]
{		\includegraphics[width=0.4\textwidth]{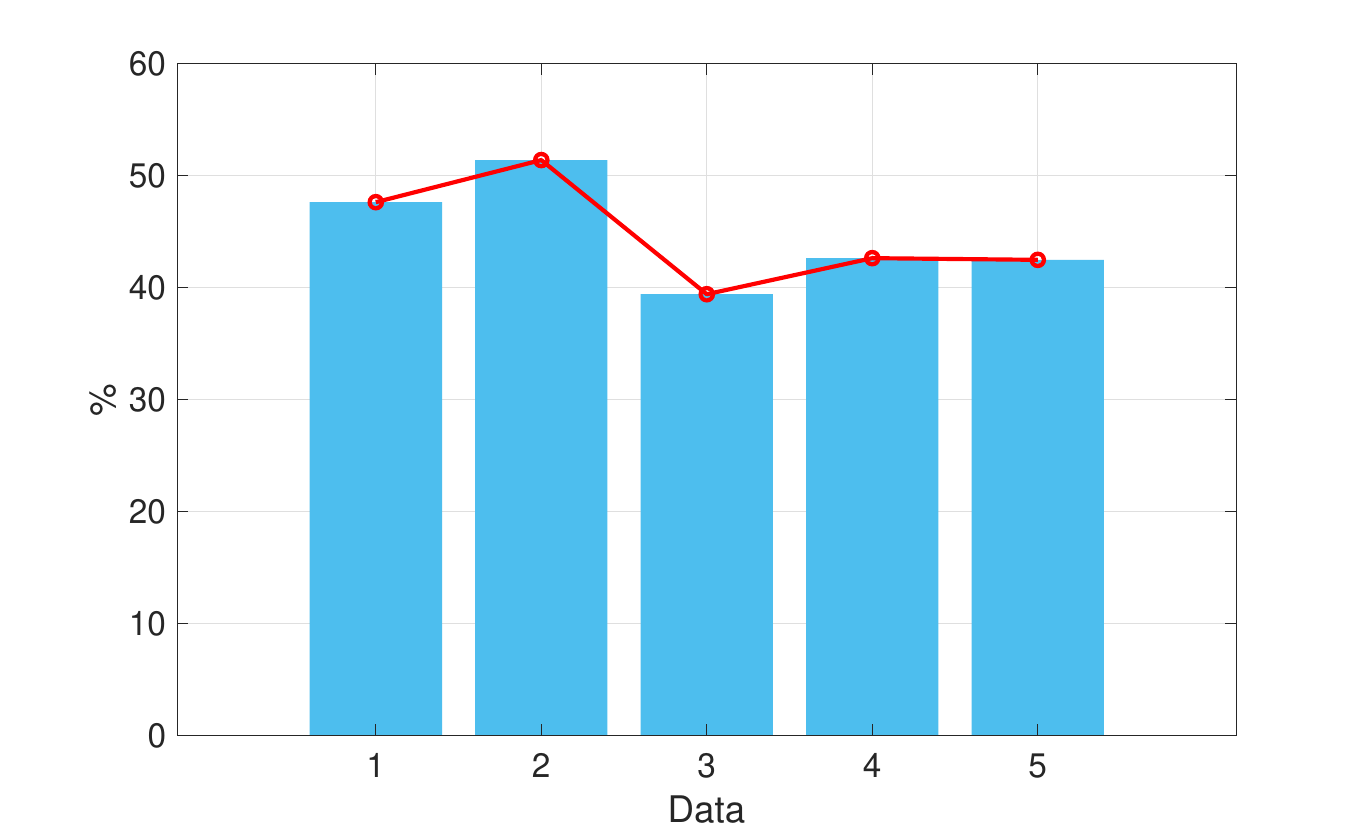}
}
\caption{The illustration of transmission time and service time reduction percentages of using the denoising-inspired prompt compression method in JPPO++ compared to the single-round compression baseline.}
\label{time}
\end{figure}
As part of the testing phase, Fig.~\ref{time} presents the transmission time and service time reduction for five randomly selected samples from the {\textit{MeetingBank-transcript}} dataset. Specifically, transmission time reduction corresponds to the time reduced by the denoising-inspired prompt compression method in JPPO++ compared to the single-round compression baseline. Service time, as illustrated in Fig.~\ref{figure1}, is defined as the sum of compression time, transmission time, and LLM inference time, capturing the end-to-end efficiency of the mobile LLM service.
Across all samples, the denoising-inspired method consistently outperforms the baseline, achieving transmission time reductions of approximately $10\%$ to $65\%$ and service time reductions of $39\%$ to $52\%$. These results confirm that the proposed JPPO++ enhances data delivery efficiency while preserving task quality. The consistent improvements across diverse prompts and varying content structures demonstrate the robustness and suitability for real-world mobile networks.

\begin{figure}[t]
\centering
\subfigure[Multi-news category]
{	\includegraphics[width=0.4\textwidth]{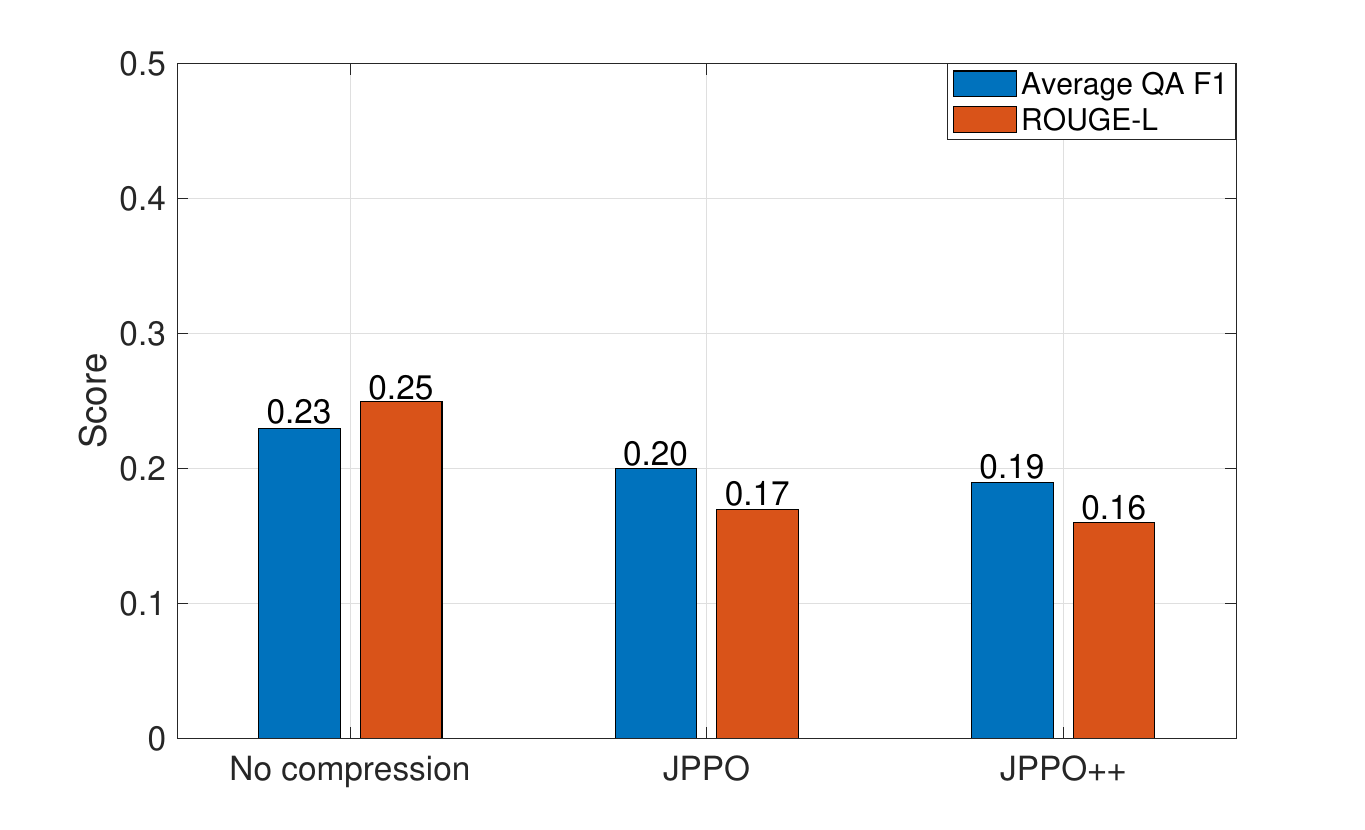}
}
\subfigure[GovReport category]
{		\includegraphics[width=0.4\textwidth]{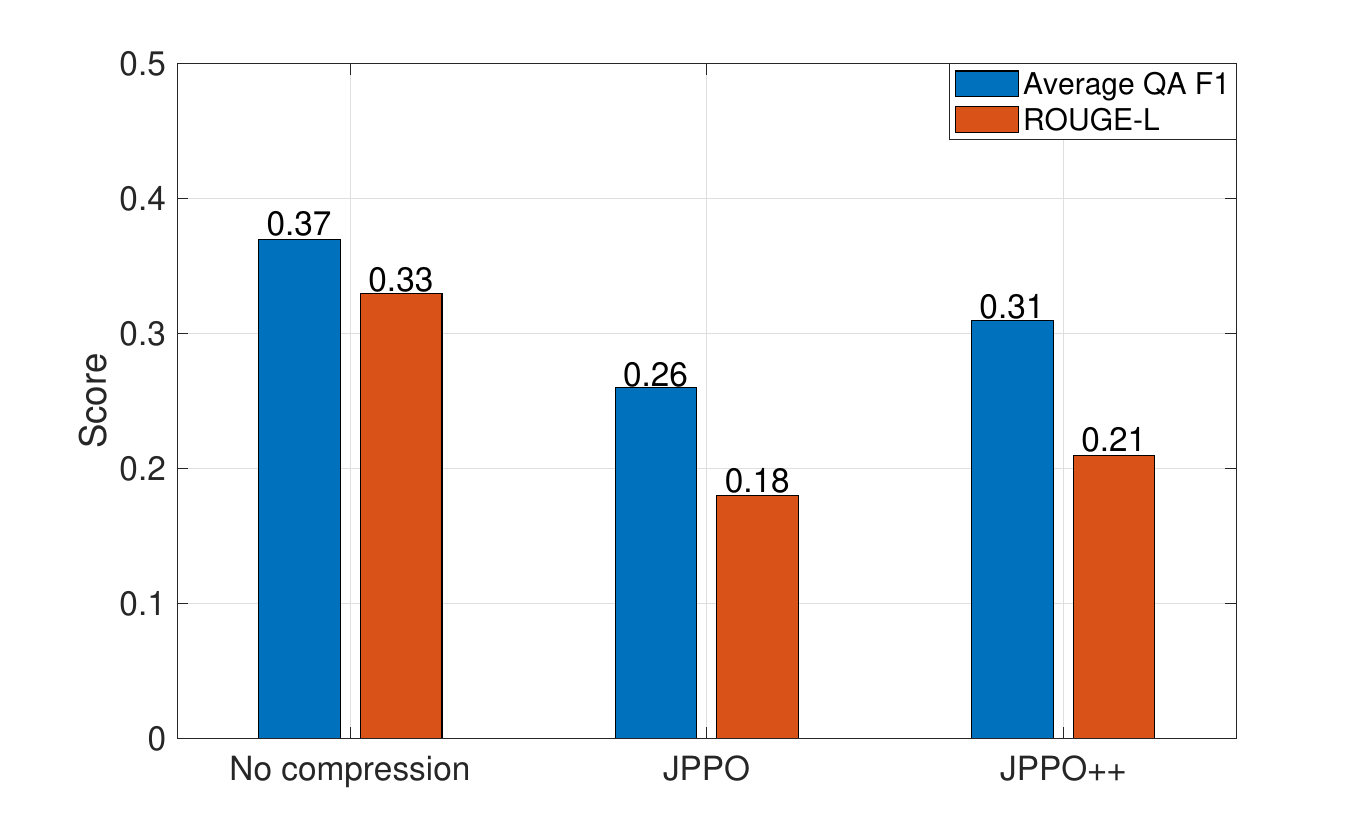}
}
\caption{The score performance when running with the \textit{LongBench} dataset.}
\label{score}
\end{figure}
Fig.~\ref{score} compares the task performance of three compression strategies, i.e., no compression, JPPO with single-round $16\times$ compression, and JPPO++ with iterative $4 \times 2\times$ compression, on the \textit{LongBench} dataset. Results are reported on both the \textit{Multi-news} and \textit{GovReport} categories using the average Question-Answering (QA) {\textit{F1}} score and ROUGE-$L$ score. The no-compression strategy achieves the highest performance across all samples. In the \textit{Multi-news} category, as shown in Fig.~\ref{score} (a), where contextual coherence is less critical, the gap between compressed and uncompressed prompts is relatively small. However, in the \textit{GovReport} category, as shown in Fig.~\ref{score} (b), which contains longer and more information-dense documents, the difference becomes more pronounced. Despite both JPPO and JPPO++ operating under the same $16\times$ compression constraint, JPPO++ consistently yields higher {\textit{F1}} and ROUGE-$L$ scores, demonstrating better retention of key information through iterative refinement.
These results highlight the practical advantage of JPPO++. It enables aggressive compression with minimal degradation of task performance. Even though uncompressed input naturally yields the best performance, JPPO++ offers a viable trade-off for resource-constrained deployments, maintaining task utility while significantly reducing prompt length.

\section{Conclusion and Future Direction}
We proposed JPPO++, a joint optimization framework that integrates denoising-inspired prompt compression with wireless transmission power control to support mobile LLM services. By leveraging a lightweight SLM for iterative prompt compression and a DRL-based policy for adapting compression ratio and power allocation, JPPO++ enabled aggressive compression while preserving response fidelity under resource constraints. The framework demonstrated stable convergence during training, and experimental results showed that it consistently reduced both service time and transmission overhead compared to single-compression baselines. Specifically, JPPO++ achieved up to $46.5\%$ service time reduction and maintained competitive performance on downstream QA and summarization tasks, even under a $16\times$ compression setting. These findings demonstrate the practical value of JPPO++ in enabling scalable and low-latency mobile LLM services.

There are several directions for further research:
\begin{itemize}
\item \textbf{Optimizing Compression Schedules :} The current JPPO++ uses a fixed denoising schedule and iteration count. Future work may treat both as decision variables. Different input types may benefit from tailored schedules, and the number of iterations directly affects computational cost and compression quality. Including these factors in the optimization can further improve performance across diverse tasks.

\item \textbf{Edge Deployment and Distributed Control:} SLMs can be deployed at edge devices to perform real-time compression before wireless transmission. Future studies may explore their interaction with LLMs in hierarchical settings. Additionally, transitioning from centralized to distributed optimization, where multiple edge nodes collaboratively manage compression and power allocation, can enhance system scalability and resilience.

\item \textbf{Personalized and Adaptive Service Provisioning:} Incorporating user-specific demands and heterogeneous network conditions into the joint optimization can enable context-aware service delivery. This is particularly relevant in autonomous systems and smart home networks, where latency and efficiency requirements vary significantly.
\end{itemize}

\bibliographystyle{IEEEtran}
\bibliography{Ref}
\end{document}